\documentclass[%
 reprint,
 amsmath,amssymb,
 aps,
]{revtex4-2}

\usepackage{amsmath}
\usepackage{bm}
\usepackage{graphicx}
\usepackage{amssymb}
\usepackage{times}
\usepackage{color}
\usepackage{float}
\usepackage{multirow}
\usepackage{subcaption}
\usepackage{braket}
\usepackage{hyperref}
\usepackage{chemformula}
\usepackage{makecell}
\usepackage{xfrac} 
\usepackage{threeparttable} 
\usepackage{siunitx}
\usepackage{booktabs}

\usepackage[
    protrusion=true,
    activate={true,nocompatibility},
    final,
    tracking=true,
    kerning=true,
    spacing=true,
    factor=1100]{microtype}
\SetTracking{encoding={*}, shape=sc}{40}

\begin{document}

\title{NV-like Defects More Common Than Four-Leaf Clovers:\\A Perspective on High-Throughput Point Defect Data}

\author{Joel Davidsson}
\email{joel.davidsson@liu.se}
\affiliation{Department of Physics, Chemistry and Biology, Link\"oping University, Link\"oping, Sweden}

\begin{abstract}

Point defects for quantum technologies is an emerging research area, with the nitrogen-vacancy (NV) center in diamond at the forefront.
However, how rare are defects with NV-like properties?
In this perspective, I highlight the results of NV-like defects across 33 different materials, revealing that they are more common than finding four-leaf clovers.
I also discuss expanding the search criteria to identify other defects relevant to quantum technologies.
Utilizing point defect databases will be instrumental in assisting researchers in discovering previously unexplored defects suitable for quantum technologies.

\end{abstract}

\maketitle


\section{Introduction}

The nitrogen-vacancy (NV) center is the original defect~\cite{Davies:PRSLA1976,gruber1997scanning,DOHERTY20131} for quantum technologies and continues to be a focus of active research and application.
It is one of the leading examples of isolated spin defects in solid-state systems, playing a crucial role in the three main areas of quantum technology: computing, communication, and sensing~\cite{Wolfowicz2021}.
The robust properties of the NV center have also inspired the development of specific criteria for finding other color centers suitable for quantum technologies.
Weber et al. outline five requirements for the defect and four for the host material~\cite{doi:10.1073/pnas.1003052107}. 

However, the NV center does have its limitations, such as optical emission in the visible range\cite{Davies:PRSLA1976}, suffering from spectral diffusion~\cite{PhysRevApplied.18.064011}, and a low Debye-Waller factor~\cite{Alkauskas_2014}.
While other defects are being explored~\cite{DiVincenzo2010,10.1063/5.0006075}, a key question remains: How rare are defects suitable for quantum technologies with NV-like properties?
The answer lies in the high-throughput data.

In order for a defect to be relevant in the realm of quantum technologies, its properties must align with the specific intended application.
Each pillar of quantum technologies requires specific defect properties, but they all revolve around a spin that can be initialized, manipulated, and read out.
Often high spin states (triplet or higher) are desirable since the high spin decouples from the electronic background of $S=\sfrac{1}{2}$ and allows for coherent spin control even at zero magnetic field~\cite{BassettAlkauskasExarhosFu}. 
To find defects that rival the NV center, in this perspective, the properties are reduced to three essential ones: defects that are stable with high spin and optical features.

While additional theoretical methods are available to model defects in greater detail~\cite{Gali}, most of these advanced techniques are currently too costly to apply in a high-throughput manner. 
Moreover, several practical challenges need to be addressed in order to realize unexplored color centers with spin.
However, this perspective does not address these challenges.

In this perspective, I show that NV-like defects are more common (conservative estimate: about 1 in 500) than finding a four-leaf clover (1 in 5,076~\cite{fourleaf}) when examining the key properties of stability, spin characteristics, and optical features.
I also discuss how to relax these criteria to find additional defects suitable for quantum technologies.
The outline of the paper is as follows:
Section~\ref{sec:adaq} introduces the Automatic Defect Analysis and Qualification (ADAQ) database.
I show that the known defects in diamond and 4H-SiC are found in the database.
I also provide an example of a predicted defect that has been realized in experiments.
Section~\ref{sec:nvlike} shows how NV-like defects are found and their trends across the periodic table.
Section~\ref{sec:search} discusses alternative ways to search for suitable defects beyond NV-like defects.
Section~\ref{sec:data} discusses how to improve the data in terms of theoretical approaches and methodologies.
Section~\ref{sec:conclusion} concludes the paper by highlighting the significant opportunity to discover unexplored defects using high-throughput data and discusses strategies to accelerate the search even further.

\section{ADAQ}
\label{sec:adaq}

There is an increase in high-throughput calculations and databases for point defects in both bulk~\cite{doi:10.1126/sciadv.adh8617,xiong2024discoverytcenterlikequantum,ivanov2023databasesemiconductorpointdefectproperties} and 2D materials~\cite{Thomas2024,Bertoldo2022,abad,Huang2023}.
In this perspective, the focus is on the ADAQ database.
The ADAQ code is the culmination of a series of publications~\cite{vacancypaper, methodologypaper, 6Hpaper, mypaper, thesis, ADAQpaper}.
In short, it is an extension of the high-throughput toolkit (\textit{httk})~\cite{armientoDatabaseDrivenHighThroughputCalculations2020}.
It uses the Vienna Ab initio Simulation Package (VASP)~\cite{VASP,VASP2} with projector augmented wave (PAW)~\cite{PAW,Kresse99} pseudopotentials.
The screening workflow relies on the Perdew, Burke, and Ernzerhof (PBE) functional~\cite{perdew1996generalized}.

ADAQ automatically generates the initial defect geometries based on symmetry. 
It can generate arbitrary-sized defect clusters, but the focus is usually on single defects (vacancy, substitutional, and interstitial) and double defects, such as substitutional-vacancy complexes.
Each geometry is processed through a screening workflow that calculates different charge and spin states, focusing initially on the neutral charge state.
The inverse participation ratio (IPR)~\cite{IPR1} is used to identify defect states in the band gap.
Additional charge ($\pm1$) and spin states are calculated when possible.
The calculation of alternative spin states is key for determining the most energetically stable spin state.
Finally, an excitation is calculated using the Delta Self-Consistent Field ($\Delta$SCF) method~\cite{cDFT}.
All calculated states, along with properties like formation energies, structural changes (such as $\Delta$Q), and optical attributes like Zero Phonon Line (ZPL) and Transition Dipole Moment (TDM)~\cite{mypaper}, are stored in a database for further filtering of relevant defects. 
Good optical features are characterized by ZPL values within an appropriate range, a large TDM, and a small $\Delta$Q.

In other high-throughput studies, the term \textit{defect} is often loosely defined and occasionally includes different charge states of the same defect geometry. 
While slight relaxations exist between the various charge states of a defect, in my opinion, the defect geometries are too similar to consider them separate defects. 
Therefore, from now on, when I refer to a \textit{defect} or count them, I specifically refer to the initial defect geometry.
For instance, in 4H-SiC, there are two symmetry-inequivalent vacancy positions; these are counted as two distinct defects or defect types.
It is crucial to note that an entry in the database is uniquely defined by the combination of defect, charge, and spin state, which adds an important layer of specificity. 
This discussion highlights the importance of standardizing the nomenclature related to defects, similar to the standardization achieved for materials with OPTIMADE~\cite{D4DD00039K}.

\subsection{Defect Hull}

Materials are ordered by thermodynamic stability using the convex hull.
Analogously, the concept of the defect hull was introduced to order the defects based on their stability~\cite{thesis,modvac}.
The defect hull consists of the point defects with the lowest formation energy for a given stoichiometry and Fermi energy.
Hence, different defects can span the hull for the same stoichiometry, such as the positive charge state of a carbon vacancy-antisite pair and the negative charge state of a silicon vacancy for the same stoichiometry (one missing silicon) in 4H-SiC~\cite{thesis}.
This result means that the defect hull can consist of complexes (double defects) and not only single defects.
In general, calculating single defects is a good starting point, but the example above highlights an important exception to include complexes.
Furthermore, the ordering within each stoichiometry of the defect hull is independent of the chemical potentials.
This attribute means that rich or poor phases do not affect any searches.
After a defect is found, the chemical potentials can be optimized to tune the absolute formation energy.
The defect hull concept has proven instrumental in finding known defects for quantum technologies.
It will also assist in identifying relevant dopants for energy-related applications, such as power converters.

\subsection{Known Defects}

Let us start by searching for known defects in diamond in the ADAQ database.
Apart from the NV center, the XV (X=Si, Ge, Sb, Pb) centers have recently attracted attention for quantum technologies~\cite{Bradac2019}.
These defects are on the defect hull with the known charge and spin states, see Table~\ref{tab:diamond}.
Two key points are worth noting.
First, the PBE functional underestimates the ZPLs by about 0.25 eV compared to experimental values~\cite{davidsson2023na}.
This underestimation primarily depends on the material, and the HSE functional yields ZPLs that closely agree with experimental results.
Second, ZPLs are missing for the Si, Ge, and Sn defects.
This result is due to one of the defect states being below the valence band, and ADAQ only calculates excitations between states within the band gap, as discussed more in Section~\ref{sec:sym}.
However, as the size of the dopant atom increases, this defect state moves into the band gap, and ADAQ captures the transition for the PbV center.
This outcome is consistent across different functionals, with the HSE functional also exhibiting the same trend~\cite{PhysRevX.8.021063}.

\begin{table}[h!]
\caption{ADAQ data for known defects in diamond. * indicates missing values, see main text for further details.}
\begin{tabular}{c|c|cc|cc}
Defect & On Defect Hull & Charge & Spin & Calc. ZPL & Exp. ZPL \\ 
& & & & [eV] & [eV] \\
\hline
    NV & yes & -1 & 1.0 & 1.700 & 1.945$\;$~\cite{Davies:PRSLA1976} \\ 
    SiV & yes & -1 & 0.5 & * & 1.680~\cite{Bradac2019}\\ 
    GeV & yes & -1 & 0.5 & * & 2.060~\cite{Bradac2019}\\ 
    SnV & yes & -1 & 0.5 & * & 2.000~\cite{Bradac2019}\\ 
    PbV  & yes & -1 & 0.5 & 2.122 & 2.384~\cite{Bradac2019} \\   
\end{tabular}
\label{tab:diamond}
\end{table}

Next, we will examine known defects in 4H-SiC.
The 4H polytype has atoms with different symmetries, meaning the same defect can exist in multiple configurations.
One of these configurations will be located on the defect hull, while the others will be positioned at a distance from the hull ($\epsilon$).
Table~\ref{tab:sic} shows the results for the divacancy and silicon vacancy.
Again, ADAQ finds the known charge and spin states, and the ZPLs are underestimated by about 0.2 eV in this case~\cite{methodologypaper, 6Hpaper, ADAQpaper}.
Notably, some configurations, such as the $kk$ configuration, show better agreement with experimental results due to error cancellation~\cite{ADAQpaper}.
For the silicon vacancy in the $k$ configuration, one of the defect states lies below the valence band, causing a missing ZPL.
In contrast to the previous discussion on defect states below the valence band, this instance can be solved by converging the calculation more thoroughly.
However, the low settings in the screening workflow are required to process the vast amount of defects effectively.

\begin{table}[h!]
\caption{ADAQ data for known defects in 4H-SiC. $\epsilon$ is the distance to the defect hull. * indicates missing values, see main text for further details.}
\begin{tabular}{cc|S[table-format=3.0]|cc|cc}
    Defect & Configuration & $\epsilon$ & Charge & Spin & Calc. ZPL & Exp. ZPL \\ 
    & & {[meV]} & & & [eV] & [eV] \\
\hline
    $\mathrm{{V_{Si}V_C}}$ & $hh$ & 20 & 0 & 1.0 & 0.992 & 1.095~\cite{Falk2013} \\ 
    $\mathrm{{V_{Si}V_C}}$ & $hk$ & 115 & 0 & 1.0 & 1.005 & 1.150~\cite{Falk2013} \\ 
    $\mathrm{{V_{Si}V_C}}$ & $kh$ & 68 & 0 & 1.0 & 0.961 & 1.119~\cite{Falk2013} \\ 
    $\mathrm{{V_{Si}V_C}}$ & $kk$ & 0 & 0 & 1.0 & 1.012 & 1.096~\cite{Falk2013} \\ 
    \hline
    $\mathrm{{V_{Si}}}$ & $h$  & 0 & -1 & 1.5 & 1.270 & 1.440~\cite{PhysRevB.62.16555} \\
    $\mathrm{{V_{Si}}}$ & $k$  & 9 & -1 & 1.5 & * & 1.353~\cite{PhysRevB.62.16555} \\  
\end{tabular}
\label{tab:sic}
\end{table}

There is a difference in formation energy between the configurations, which is reflected in the distance to the defect hull.
For the divacancy, the $kk$ configuration is on the defect hull, but the other configurations are also observed experimentally.
The $hk$ configuration is 115 meV higher in energy than the $kk$ configuration, indicating that some defects above the defect hull can be observed. 
For now, let us set the distance to the hull to 30 meV and discuss this further in Section~\ref{sec:hull}.

I would like to make two final remarks about the results in 4H-SiC.
First, according to the ADAQ database, the carbon vacancy-antisite pair (CAV) is not related to the AB lines, which is also supported by higher-order calculations (GW)~\cite{cavpaper}.
Instead, the CAV is predicted to emit close to the telecom bands~\cite{cavpaper}.
Last, while searching for the silicon vacancy in the ADAQ database, additional defects with the same spin and similar ZPLs were also found.
These defects are the modified silicon vacancies, which consist of a silicon vacancy with a carbon antisite as a second nearest neighbor.
These defects were identified through experiments, demonstrating the predictive power of high-throughput methods~\cite{modvac}.

\section{NV-like Defects}
\label{sec:nvlike}

This section highlights various host materials for quantum technologies and their number of defects, outlines criteria for NV-like defects, and details several selected defects discovered through the high-throughput approach.

\subsection{Hosts}

Research continues on defects in traditional materials for quantum technologies like diamond and SiC.
ADAQ has processed 8,450 single and double defects in diamond\footnote{Originally there were 21,607 defects, but a bug created duplicates. After removing these duplicates, 8,450 unique defects remained.}~\cite{davidsson2023na} and 60,081 single and double defects in 4H-SiC~\cite{thesis, modvac, clvpaper}.
While both diamond and SiC are great hosts, other materials with longer coherence times have also been suggested.
Kanai et al. ranked the materials in the Materials Project~\cite{10.1063/1.4812323} based on their coherence time and found that oxides are at the top~\cite{kanai}.
Notably, calcium oxide (CaO) ranks third among all computed materials (first if materials with d- and f-elements are excluded).
Magnesium oxide (MgO) shares a similar ranking to diamond.
Single and double defects were calculated in these materials, resulting in 9,077 defects in CaO~\cite{davidsson2024discovery} and 2,917 defects in MgO~\cite{somjit2024nvcentermagnesiumoxide}\footnote{More defects in CaO due to inclusion of complexes up to fourth nearest neighbors.}.
Additionally, Ferrenti et al. reduced the Materials Project database~\cite{10.1063/1.4812323} to 541 materials with suitable properties (stable, low number of nuclear spin isotopes, band gap larger than 1.1 eV) for hosting quantum defects~\cite{Ferrenti2020}.
Groppfeldt et al. further reduced this list and calculated 17,710 single defects in 29 materials~\cite{oscarmaster}.
All materials were doped with s- and p-elements from H to Bi.
In total, about 100,000 defects (initial defect geometries) have been processed by ADAQ, which can now be filtered for quantum technology applications.

\subsection{Search Criteria}

In this perspective, NV-like defects are found with the following search criteria.
The defect should be:
\begin{itemize}
    \item stable within 30 meV of the defect hull ($\epsilon<30$ meV),
    \item with a spin of 1 or greater ($S\ge1$),
    \item and an optical signal (ZPL$>$0.5 eV and TDM$>$3 Debye).
\end{itemize}
In Section~\ref{sec:search}, I discuss what happens if we relax these search criteria.

\subsection{Defects}

\begin{figure*}[t]
  \includegraphics[width=\textwidth]{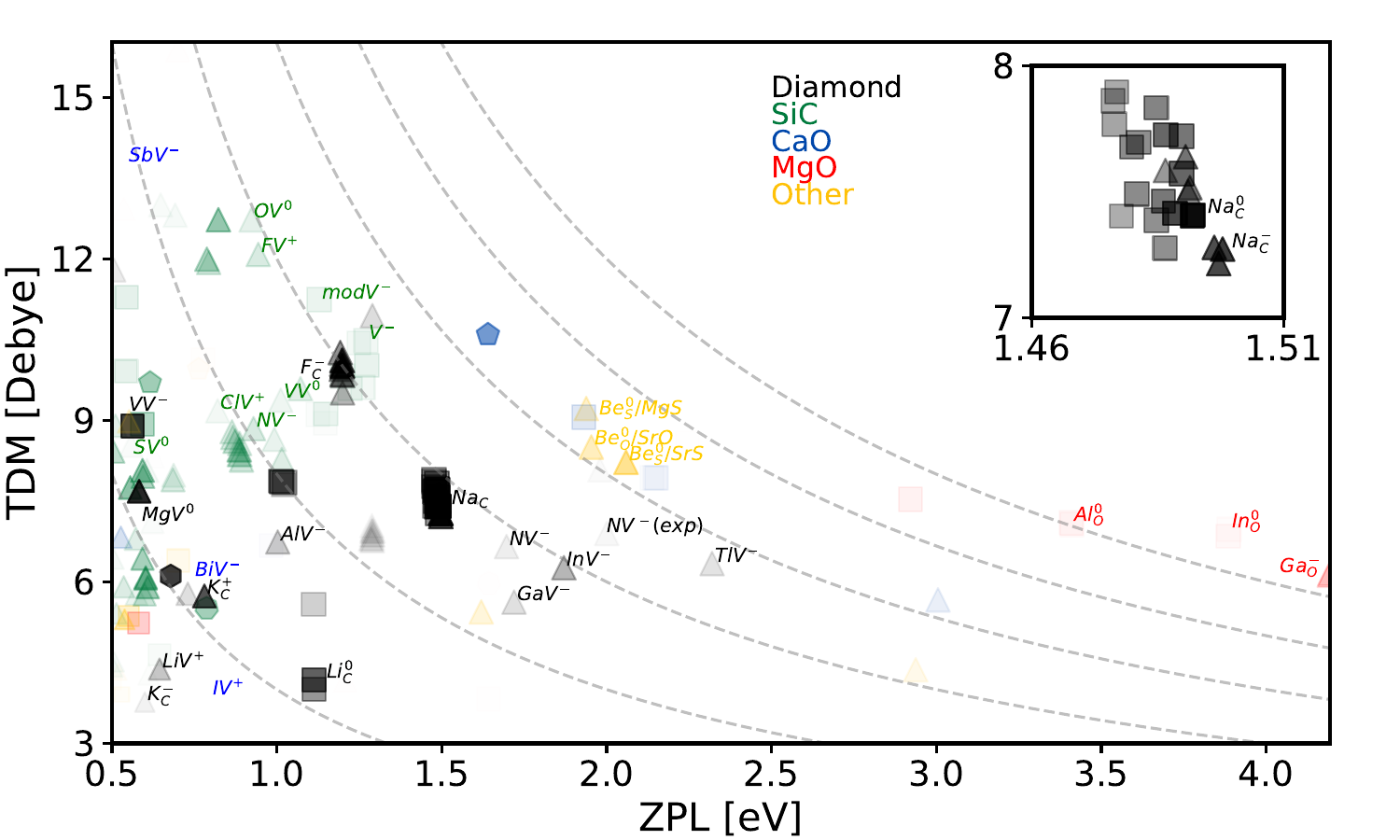}
	\caption{ADAQ (PBE) results for defects that fulfill the search criteria. Near the center, a point shows the experimental characteristics of the NV center. The corners of the marker indicate the stable ground state; for instance, a triangle represents a triplet ground state. The size of each marker corresponds to the radiative lifetime (larger markers correspond to faster rates), while the opacity reflects the Debye-Waller factor. $V$ represents vacancy.  Guidelines follow 1/ZPL curves~\cite{thesis}. The inset shows Frenkel pairs ($\mathrm{Int_{Na}V_C}$) that relaxed into the Na substitutional ($\mathrm{Na_C}$).}
	\label{fig:overview} 
\end{figure*}

This search yields 287 NV-like defect entries across 33 materials.
These are presented in Figure~\ref{fig:overview}.
It is worth noting that some of these results include duplicates (when removed, 180 defects remain), as the initial structures can relax into the same configuration.
Many of the filtered defects are challenging to realize because they involve second-nearest neighbor complexes. 
Finally, we have not ranked these defects to determine whether they are superior to the NV center, although 111 have a Debye-Waller factor (calculated using the one-phonon approximation) larger than 3\%.
Nonetheless, while NV-like defects are rare, they are not exceptionally rare. 
The data indicate that it is easier to find an NV-like defect~\footnote{ratio of NV-like defects lies somewhere from 1 in 600 (180 NV-like defects to 100,000 processed defects) to 1 in 60 (compared to ca 10,000 stable defects on the defect hull).} than a four-leaf clover~\cite{fourleaf}.
Several defects stand out and have been selected for further characterization, with results in the Table~\ref{tab:nvlike}.

\newcolumntype{P}[1]{>{\centering\arraybackslash}p{#1}}
\newcommand\sep{1.5pt}

\begin{table*}[t]
\caption{Selected NV-like defects from the ADAQ database. Found using the NV-like search criteria listed in the text. V indicates vacancy. Added the NV center as a reference.}
\begin{tabular} {lc|rP{0.85cm}|P{1cm}S[table-format=2.2]c|l}
Host & Defect & Charge & Spin & ZPL & {TDM} & $\Delta$Q & Note \\
& (configuration) & & & [eV] & {[debye]} & [amu$^{1/2}$\AA] &  \\
\hline
\multirow{10}{*}{diamond} & \multirow{2}{*}{NV} & \multirow{2}{*}{-1} & \multirow{2}{*}{1.0} & 1.70 & 6.66 & 0.56 & ADAQ results~\cite{davidsson2023na} \\
& & & & 2.00 & 6.92 & 0.71 & HSE results~\cite{davidsson2024discovery, Alkauskas_2014} \\[\sep]
& $\mathrm{Li_CV_C}$ & +1 & 1.0 & 0.64 & 4.38 & 0.41 & ADAQ results~\cite{davidsson2023na} \\
& $\mathrm{Li_C}$ & 0 & 1.5 & 1.11 & 4.17 & 0.21 & ADAQ results~\cite{davidsson2023na} \\
& $\mathrm{K_C}$ & -1 & 1.0 & 0.59 & 3.77 & 0.96 & ADAQ results~\cite{davidsson2023na} \\
& $\mathrm{K_C}$ & +1 & 1.0 & 0.78 & 5.74 & 0.24 & ADAQ results~\cite{davidsson2023na} \\[\sep]
& \multirow{2}{*}{$\mathrm{Na_C}$} & \multirow{2}{*}{0} & \multirow{2}{*}{1.5} & 1.49 & 7.41 & 0.16 & ADAQ results~\cite{davidsson2023na} \\
& & & & 1.59 & 6.8 & 0.52 & HSE results~\cite{davidsson2023na} \\[\sep]
& \multirow{2}{*}{$\mathrm{Na_C}$} & \multirow{2}{*}{-1} & \multirow{2}{*}{1.0} & 1.50 & 7.27 & 0.15 & ADAQ results~\cite{davidsson2023na} \\
& & & & 1.68 & 6.6 & 0.69 & HSE results~\cite{davidsson2023na} \\
\hline

\multirow{9}{*}{4H-SiC} & $\mathrm{N_CV_{Si}}(hh)$ & -1 & 1.0 & 0.93 & 8.85 & 0.59 & ADAQ results~\cite{clvpaper} \\[\sep]
& \multirow{2}{*}{$\mathrm{O_CV_{Si}}(hh)$} & \multirow{2}{*}{0} & \multirow{2}{*}{1.0} & 1.07 & 9.59 & 0.63 & ADAQ results~\cite{clvpaper} \\
& & & & 1.24 & 10.42 & 0.79 & HSE results~\cite{abbas2025theoreticalcharacterizationnvlikedefects}\\[\sep]
& \multirow{2}{*}{$\mathrm{S_CV_{Si}}(hh)$} & \multirow{2}{*}{0} & \multirow{2}{*}{1.0} & 0.61 & 8.07 & 0.86 & ADAQ results~\cite{clvpaper} \\
& & & & 0.67 & 9.42 & 1.02 & HSE results~\cite{abbas2025theoreticalcharacterizationnvlikedefects} \\[\sep]
& \multirow{2}{*}{$\mathrm{F_CV_{Si}}$\begin{tabular}{@{}c@{}}$(hk)$\\$(hh)$\end{tabular}} & \multirow{2}{*}{+1} & \multirow{2}{*}{1.0} & 0.94 & 12.08 & 0.53 & ADAQ results~\cite{clvpaper} \\
& & & & 1.26 & 10.50 & 1.08 & HSE results~\cite{abbas2025theoreticalcharacterizationnvlikedefects} \\[\sep]
& \multirow{2}{*}{$\mathrm{Cl_CV_{Si}}(hh)$} & \multirow{2}{*}{+1} & \multirow{2}{*}{1.0} & 0.82 & 9.18 & 0.83 & ADAQ results~\cite{clvpaper} \\
& & & & 0.93 & 9.99 & 0.96 & HSE results~\cite{clvpaper} \\
\hline

\multirow{6}{*}{CaO} & \multirow{2}{*}{$\mathrm{Sb_{Ca}V_O}$} & \multirow{2}{*}{-1} & \multirow{2}{*}{1.0} & 0.54 & 13.8 & 4.39 & ADAQ results~\cite{davidsson2024discovery} \\
& & & & 0.53 & 34.7 & 5.10 & HSE results~\cite{davidsson2024discovery} \\[\sep]
& \multirow{2}{*}{$\mathrm{Bi_{Ca}V_O}$}  & \multirow{2}{*}{-1} & \multirow{2}{*}{1.0} & 0.74 & 5.99 & 5.05 & ADAQ results~\cite{davidsson2024discovery} \\
& & & & 0.76 & 10.8 & 6.22 & HSE results~\cite{davidsson2024discovery} \\[\sep]
& \multirow{2}{*}{$\mathrm{I_{Ca}V_O}$}  & \multirow{2}{*}{+1} & \multirow{2}{*}{1.0} & 0.79 & 3.72 & 3.20 & ADAQ results~\cite{davidsson2024discovery} \\
& & & & 0.78 & 41.8 & 5.49 & HSE results~\cite{davidsson2024discovery} \\
\hline

\multirow{2}{*}{MgO} & \multirow{2}{*}{$\mathrm{Int_{N}V_{Mg}}$} & \multirow{2}{*}{-1} & \multirow{2}{*}{1.0} & 2.19 & $^\ast$0.66 & 1.95 & ADAQ results~\cite{somjit2024nvcentermagnesiumoxide} \\
& & & & 3.19 & 0.17 & 3.37 & Hybrid TDDFT results~\cite{somjit2024nvcentermagnesiumoxide} \\
\hline

SrS & $\mathrm{Be_{S}}$ & 0 & 1.0 & 2.06 & 8.22 & 0.55 & ADAQ results~\cite{oscarmaster} \\
MgS & $\mathrm{Be_{S}}$ & 0 & 1.0 &  1.94 & 9.22 & 0.65 & ADAQ results~\cite{oscarmaster} \\
SrO & $\mathrm{Be_{O}}$ & 0 & 1.0 & 1.95 & 8.51 & 0.67 & ADAQ results~\cite{oscarmaster} \\
\end{tabular}
\label{tab:nvlike}

\smallskip\footnotesize\centering 
$\ast$No TDM restriction in search
\end{table*}

Apart from the NV center in diamond, other defects exhibit similar properties.
Among the various XV defects are the group 13 vacancy clusters, which have been suggested previously~\cite{PhysRevB.102.195206}, but also the LiV defect in the positive charge state.
Moreover, the group 1 substitutionals also have high spins (spin-\sfrac{3}{2} in neutral charge state) and great optical features.
Here, Davidsson et al. focused on the $\mathrm{Na_C}$ that have NV-like properties in two charge states, highlighting that one defect can have multiple charge and spin states with NV-like properties.
Both charge states of $\mathrm{Na_C}$ has a lower ZPL and lower $\Delta$Q compared to the NV center~\cite{davidsson2023na}.

In 4H-SiC, the nitrogen-vacancy (NV) center is also present, as previously established~\cite{Bardeleben_Cantin_2017}.
However, the high-throughput search identified four additional XV defects that are listed in Table~\ref{tab:nvlike}.
As the group is changed, the charge state with NV-like properties also changes.
Specifically, O and S defects in the neutral charge state have NV-like properties, whereas F and Cl need to be in the positive charge state.
The oxygen vacancy (OV) defect was presented by a separate research group in close connection with the ADAQ results~\cite{10.1063/5.0169147}.
Using a high-throughput method enabled the full characterization of the defect space and the identification of the full suite of NV-like defects in 4H-SiC. 
Bulancea-Lindvall et al. focused on the ClV defect in more detail, as it is predicted to have a ZPL in the telecom band~\cite{clvpaper}.

So far, we have examined conventional covalent materials, such as diamond and SiC, concerning color centers with spin.
Now, we shift to oxides, which are predicted to have longer coherence times, starting with CaO.
Again, a class of defects with NV-like properties was found, even though this material is more ionic than previous materials.
The XV defects are mostly found in group 15 of the periodic table, where P, As, Sb, and Bi dopants in the negative charge state exhibit NV-like properties.
However, since P and As have two local minima, they are excluded from our current discussion.
In group 17, the I dopant in the positive charge state also has NV-like properties.
In group 16, no NV-like defects have been observed due to the negative U behavior of the XV defects, \textit{i.e.}, the neutral charge state is unstable.
Furthermore, the BiV defect also gives rise to a clock transition, where the avoided spin states increase the coherence time by eliminating the first-order decoherence term~\cite{davidsson2024discovery}.
This result raises an open question: How common are clock transitions among color centers?

MgO also has the same defect suite as CaO.
However, it also has a unique NV-like defect, the nitrogen interstitial combined with a magnesium vacancy~\cite{somjit2024nvcentermagnesiumoxide}.
In this defect, the N binds to O, forming a dimer.
This example illustrates the importance of considering more complicated defect structures, as the single defect ($\mathrm{N_{Mg}}$) has an energy approximately 1.5 eV higher.
Although oxides are predicted to have better coherence times, other properties might be worse.
A notable difference in both MgO and CaO, when compared to diamond and SiC, is the larger $\Delta$Q.
The ionic nature, materials with lower stiffness, and defects containing vacancies contribute to the larger $\Delta$Q, which will limit the Debye-Waller factor.
Nevertheless, there may be potential for material engineering to enhance the Debye-Waller factor.
Alternatively, other readout mechanisms may be more suitable for these defects, as discussed in Section~\ref{sec:readout}.

Finally, Groppfeldt et al. studied single defects in 29 host materials using ADAQ and found 13 NV-like single defects with Debye-Waller factor $>$ 1\%.
Among these defects, 3 were Be substitutionals ($\mathrm{Be_{S}}$ in SrS, $\mathrm{Be_{S}}$ in MgS, $\mathrm{Be_{O}}$ in SrO)~\cite{oscarmaster}.
These defects are also found in more ionic materials, but still have reasonable $\Delta$Q comparable to those in SiC and diamond.
This finding emphasizes that alternative materials can host NV-like defects with properties on par with the NV center in diamond.

\subsection{Trends}

\begin{figure*}[t]
  \includegraphics[width=\textwidth]{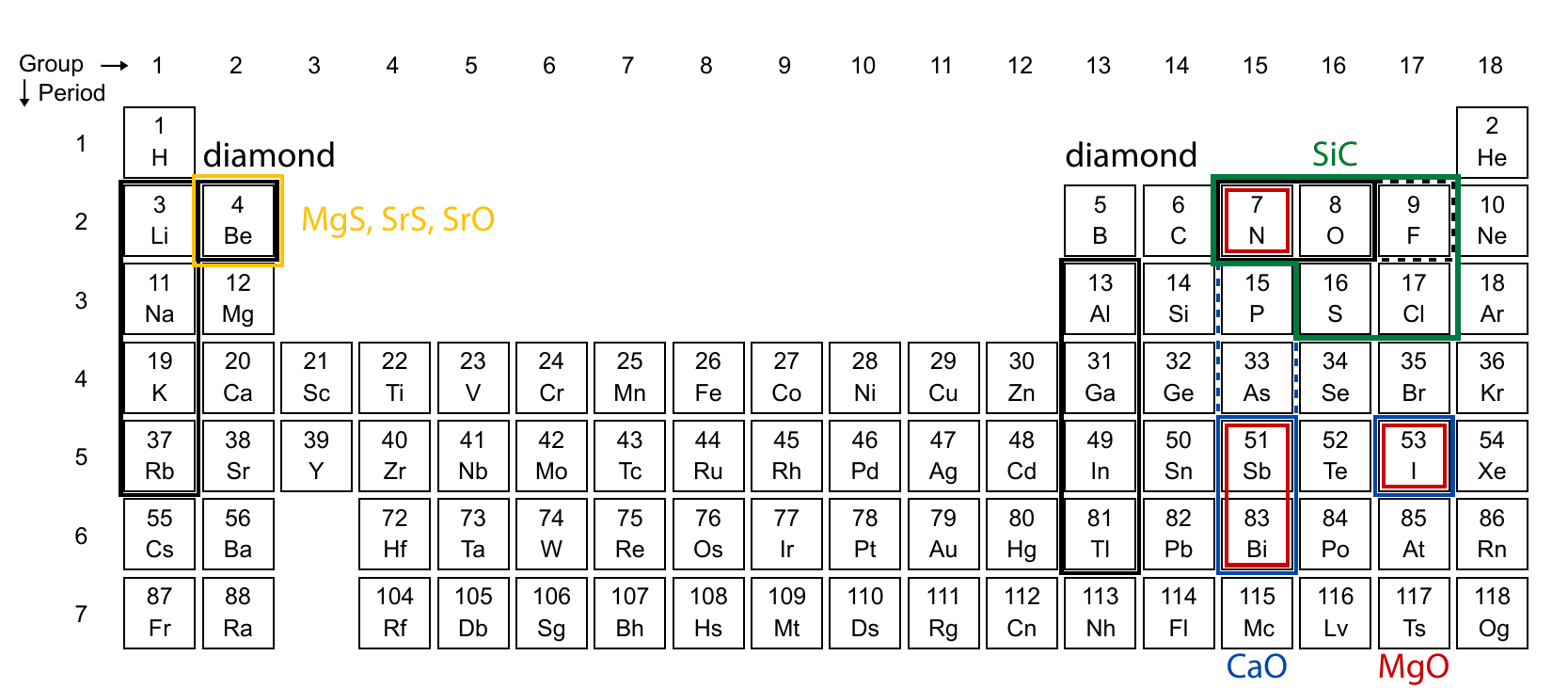}
	\caption{Trends of NV-like defects across the periodic table are represented as follows: defects in diamond are marked in black, defects in SiC in green, CaO in blue, and MgO in red. Defects in MgS, SrS, and SrO are marked in yellow. Dashed lines represent defects with uncertain trends.}
	\label{fig:trends} 
\end{figure*}

Figure~\ref{fig:trends} illustrates trends in high-throughput data for NV-like defects.
First and foremost, the NV center is found in various materials, including diamond, SiC, and MgO.
Although the form and stoichiometry of the NV center in MgO differ from those in diamond, it still involves a nitrogen atom and a vacancy.
Furthermore, in certain instances, it is possible to replace nitrogen, as demonstrated by the XV defects (X=O, S, F, Cl) in SiC, and still retain the NV-like properties.
Similar trends can be observed in diamond, although they are not as extensive.
The neutral OV also exists in diamond, as predicted in earlier works~\cite{PhysRevB.72.035214,10.1063/1.4892654}.
In addition, the ADAQ data suggests that the negative charge state of OV has spin-$\sfrac{3}{2}$.
Moreover, the FV in diamond follows the trend with a spin-1 ground state in the positive charge state.
It also has a spin-2 ground state in the negative charge state, although additional data is needed to confirm this.

These defects are primarily found within the same part of the periodic table, mainly along the rows.
There are also trends within the groups; for example, the group 13 vacancy clusters in diamond are one example of predicted defects~\cite{PhysRevB.102.195206}.
Recent DFT modeling of these defects further reinforces the notion that AlV, GaV, and InV are ready for experimental verification~\cite{GOSS2024110811}.
Additionally, the XV defects (X=Sb, Bi, I) in CaO and MgO also have NV-like properties.

These examples are taken from the p-block of the periodic table, where the defects are primarily substitutional-vacancy complexes.
However, there are also examples in the s-block of substitutional defects. 
For example, group 1 substitutionals (Li, N, K, and Rb) in diamond also have NV-like properties~\cite{davidsson2023na}.
Additionally, Beryllium, which belongs to group 2, is found in materials such as SrS, MgS, and SrO~\cite{oscarmaster}.
This result shows that defect trends exist across different materials. 
As long as the defect states remain within the band gap and the material does not change significantly, the properties of the impurity are preserved.
The results depicted in Figure~\ref{fig:trends} demonstrate the presence of NV-like defects throughout the periodic table.
Note that the d-block is currently unrepresented, as no defects involving d-elements have been calculated.
However, based on the trends observed in the s- and p-blocks, it is probable that many NV-like defects also exist in the d-block.

\section{Other Searches}
\label{sec:search}

How similar does a defect need to be to the NV center in order to be considered useful? 
It is possible to relax certain criteria. 
In this section, we will examine each criterion used to identify NV-like defects to explore other potential ways to discover interesting defects.

\subsection{Defect hull}
\label{sec:hull}

The stability of defects is a crucial factor to consider.
The exact distance to the defect hull remains a topic of discussion.
For instance, as indicated for the divacancy in Table~\ref{tab:sic}, the distance can exceed 0.1 eV, yet it can still be realized experimentally.
Therefore, a small buffer from the defect hull is necessary.
Another reason for having a buffer is that the spin states may be close in energy, and higher-order calculations could shift these states to favor one over the others.
A buffer of 0.1 eV may be sufficient for the PBE data, and it is discussed more in Section~\ref{sec:data}.

There is also the remote possibility that defects are metastable.
They could get stuck in the local geometry and locked within the crystal.
In this context, diffusion barriers play a crucial role, they are also discussed in Section~\ref{sec:data}.
Increasing the distance to the defect hull to 1 eV or greater makes it possible to include metastable defects.
However, additional checks must be performed to ensure that no other defects are closer to the defect hull, neither in terms of lower-lying spin states of the same defect nor a relaxation pathway that could transform the metastable defect to another more stable defect.
Metastable defects for quantum applications are highly unlikely, but further studies are needed.

Finally, the defect hull criterion requires further refinement. 
The current criteria, which emphasize stability for any Fermi level, may be overly broad.
Improving this approach involves considering the actual Fermi level and the feasibility of doping to achieve it.
Additionally, it should also assess whether the defect remains stable over a wide or narrow range of Fermi levels. 
Updating this criterion could help narrow down the list of defects more efficiently.

\subsection{Spin}

So far, the search criterion has been for spin-1 or higher.
The ratio of triplet to singlet and doublet ground states depends on the host material.
Traditional viewpoints emphasize symmetry.
If the symmetry of the host material drops below space group 74, the crystal no longer has any degenerate states. 
Consequently, the availability of triplet ground states should depend on the symmetry of the host materials.
Generally, high-symmetry materials should have a higher incidence of high-spin ground states. 
However, while it is less likely to find high-spin ground states in low-symmetry crystals, it is not impossible. 
Without degenerate levels, high-spin states occur when the exchange coupling between electrons in different defect states exceeds the single-particle energy difference between those states~\cite{BassettAlkauskasExarhosFu}.
For example, the silicon vacancy in SiC has spin-$\sfrac{3}{2}$ even though the crystal has lower symmetry without triple degenerate states.
Furthermore, unexpectedly many spin defects with high spin were found in systems with low symmetry~\cite{oscarmaster}.
Additionally, low-symmetry environments can create defects with anisotropic g-factors that offer faster driving speeds and lower decoherence~\cite{Wolfowicz2021}.

While triplet or higher spin ground states are preferable, a doublet ground state may be sufficient.
For instance, the XV defects--—where X represents Si, Ge, Sn, or Pb in diamond—--have received significant attention, and they have a doublet ground state in the negative charge state~\cite{Bradac2019}.
If we broaden the spin criterion to $S\ge\sfrac{1}{2}$, the number of defects increases dramatically since about half of all defects in a material have a doublet ground state (about 46\% in 4H-SiC~\cite{thesis}) regardless of the symmetry of the host.

\subsection{Readout}
\label{sec:readout}

Optical readout is generally preferred in many applications~\cite{Wolfowicz2021}.
For effective optical readout, it is important to have a bright ZPL and a large Debye-Waller factor.
Additionally, considering relaxation processes in the excited state is crucial, both for estimating the Debye-Waller factor and because certain defects exhibit significant relaxation in the excited state.
For example, $\mathrm{O_{Si}}$ in 4H-SiC has an absorption of 1.76 eV, whereas the ZPL is around 0.8 eV.

However, electrical or mechanical readout offers alternative control possibilities, as demonstrated for the NV center in diamond~\cite{7478018}.
Hence, the requirements of a bright ZPL and a large Debye-Waller factor may not be necessary in all cases; it depends on the application.
In quantum communication, where flying qubits are essential, optical readout is indeed important.
However, for quantum computing or sensing, alternative readout protocols may be more beneficial.
Therefore, it is necessary to update databases to filter for the relevant properties that help identify defects suitable for electrical or mechanical readout.

\section{Improving the Data}
\label{sec:data}

Searches have been applied to the screening data, which is produced with the PBE functional.
The results obtained from this filtering process are only as reliable as the data itself.
The PBE functional has known weaknesses, including smaller band gaps and possible incorrect ordering of spin states.
There is, of course, a risk that we overlook interesting systems due to these errors.
This challenge is common in all high-throughput approaches, and the goal is to mitigate these errors.
Despite utilizing relatively low convergence settings and the PBE functional, intriguing defects have still been identified in the data.
It is essential to validate every prediction derived from the screening data using higher-order methods.

With this context, let us discuss how to improve the screening data.
I see three possible avenues for enhancement:
i) Implementing a HSE screening workflow that will improve the band gap and relative spin state energies.
However, since it is more expensive, it is necessary to rely on geometries that are relaxed with the PBE functional.
ii) Expanding the database with additional properties is a logical next step. 
Among the top priorities are spin-related properties, such as zero-field splitting and hyperfine coupling parameters. 
Another important aspect to consider is the various rates for spin relaxation pathways. 
It would be beneficial to have a rapid method for determining whether a defect has an optical cycle and if its spin can be initialized. 
Currently, existing methods are too expensive to implement in a high-throughput workflow.
iii) Instead of pursuing more accurate functionals and methods, we could focus on developing faster methods at the PBE level, which would also enhance the screening data.
There are many defect systems left to explore.
If we had faster methods, we could evaluate more defects, larger complexes, more charge states, and more excitations.
Additionally, we should investigate co-doping strategies, as current defect analyses typically involve only one external dopant, rather than mixtures.
Combining different dopants may be necessary to realize specific defects in the desired charge state.
Furthermore, calculating energy barriers using nudge-elastic band methods and addressing defects in stacking faults~\cite{stackingpaper} also requires faster methods.
Using the SCAN functional could improve performance in both areas, as demonstrated for NV-like defects in 4H-SiC, where many properties align more closely with the HSE results while maintaining a similar computational cost to the PBE functional~\cite{abbas2025theoreticalcharacterizationnvlikedefects}.

\subsection{Symmetry Analysis}
\label{sec:sym}

In parallel to updating the computational methods, there is also a need to update the algorithms used for identifying defects and selecting transitions.
Currently, the ADAQ screening workflow selects only one excitation per defect, potentially overlooking important excitations between defect states within the band gap or selecting forbidden transitions.
One potential solution is to incorporate symmetry analysis.
The recently published code, ADAQ-SYM, calculates the character of defect states and determines the selection rules for any given defect~\cite{STENLUND2025109468}.
This approach was verified on the SiV in diamond, where the correct excitation was identified.
The intention is to apply ADAQ-SYM to the ADAQ database to search for additional ZPLs that may have been previously overlooked.

The selection rules are also valuable for gaining a deeper understanding of the defects, and may allow for the optical selection of which defect to excite.
For the divacancy and NV center in SiC, it is possible to avoid exciting the high-symmetry configurations, which leads to an increase in the ODMR contrast for the low-symmetry configurations~\cite{shafizadeh2023selection}.
However, this possibility is not available for all defects and depends on the structure and symmetry of the defect.

Understanding the symmetry properties is not only crucial for determining transitions but also for gaining a deeper insight into the defects themselves.
Moreover, the symmetry knowledge is a step towards generating many-body diagrams and exploring strain deformations.

\section{Conclusion}
\label{sec:conclusion}

High-throughput data have identified and predicted defects for quantum technologies, with many unexplored defects that may rival the NV center in diamond.
Numerous defects are awaiting further theoretical investigation and experimental realization.
On the theoretical side, advancements in theoretical characterization and more data are needed to guide experimental efforts. 
Currently, the ADAQ database allows users to filter defects but lacks recipes for creating them. 
Improved methods could provide guidelines on sample preparation, doping concentrations, and annealing conditions for generating or optimizing specific defects. 
On the experimental side, I would like to see a structured approach for generating point defects, ideally with an automated setup that implants, anneals, and measures multiple samples while documenting each step of the process in a public database.

We are at the beginning of the research field dedicated to point defects for quantum technologies, which has the potential to become the dominant quantum technology.
There will not be a single defect system for all applications; rather, multiple defect systems tailored for specific applications.
Although a significant amount of work is required to bring defects to the same level as the NV center, existing databases currently provide many promising defects for further study.
The most significant defects may still be undiscovered, and the choice of host material remains undecided.
However, with more systematic data collection and analysis, we will come closer to identifying the optimal defects, and I expect that more NV-like defects will be identified in the future.

\section*{Acknowledgments}

The author thanks Oscar Bulancea Lindvall, Rickard Armiento, and Igor Abrikosov for valuable discussions and insights on the manuscript. 

The author acknowledges funding from the Swedish Research Council (VR) Grant No. 2022-00276 and the Knut and Alice Wallenberg Foundation through the WBSQD2 project (Grant No. 2018.0071).

Most of the ADAQ computations discussed in this paper were enabled by resources provided by the National Academic Infrastructure for Supercomputing in Sweden (NAISS), partially funded by the Swedish Research Council through grant agreement no. 2022-06725.

\section*{Author Declaration}

The author has no conflicts to disclose.

\section*{Data Availability}

The data that support the findings of this study are openly available in the ADAQ database (\url{https://defects.anyterial.se/}).

\bibliography{references}

\begin{thebibliography}{62}%
\makeatletter
\providecommand \@ifxundefined [1]{%
 \@ifx{#1\undefined}
}%
\providecommand \@ifnum [1]{%
 \ifnum #1\expandafter \@firstoftwo
 \else \expandafter \@secondoftwo
 \fi
}%
\providecommand \@ifx [1]{%
 \ifx #1\expandafter \@firstoftwo
 \else \expandafter \@secondoftwo
 \fi
}%
\providecommand \natexlab [1]{#1}%
\providecommand \enquote  [1]{``#1''}%
\providecommand \bibnamefont  [1]{#1}%
\providecommand \bibfnamefont [1]{#1}%
\providecommand \citenamefont [1]{#1}%
\providecommand \href@noop [0]{\@secondoftwo}%
\providecommand \href [0]{\begingroup \@sanitize@url \@href}%
\providecommand \@href[1]{\@@startlink{#1}\@@href}%
\providecommand \@@href[1]{\endgroup#1\@@endlink}%
\providecommand \@sanitize@url [0]{\catcode `\\12\catcode `\$12\catcode `\&12\catcode `\#12\catcode `\^12\catcode `\_12\catcode `\%12\relax}%
\providecommand \@@startlink[1]{}%
\providecommand \@@endlink[0]{}%
\providecommand \url  [0]{\begingroup\@sanitize@url \@url }%
\providecommand \@url [1]{\endgroup\@href {#1}{\urlprefix }}%
\providecommand \urlprefix  [0]{URL }%
\providecommand \Eprint [0]{\href }%
\providecommand \doibase [0]{https://doi.org/}%
\providecommand \selectlanguage [0]{\@gobble}%
\providecommand \bibinfo  [0]{\@secondoftwo}%
\providecommand \bibfield  [0]{\@secondoftwo}%
\providecommand \translation [1]{[#1]}%
\providecommand \BibitemOpen [0]{}%
\providecommand \bibitemStop [0]{}%
\providecommand \bibitemNoStop [0]{.\EOS\space}%
\providecommand \EOS [0]{\spacefactor3000\relax}%
\providecommand \BibitemShut  [1]{\csname bibitem#1\endcsname}%
\let\auto@bib@innerbib\@empty
\bibitem [{\citenamefont {Davies}\ and\ \citenamefont {Hamer}(1976)}]{Davies:PRSLA1976}%
  \BibitemOpen
  \bibfield  {author} {\bibinfo {author} {\bibfnamefont {G.}~\bibnamefont {Davies}}\ and\ \bibinfo {author} {\bibfnamefont {M.~F.}\ \bibnamefont {Hamer}},\ }\bibfield  {title} {\bibinfo {title} {Optical studies of the 1.945 ev vibronic band in diamond},\ }\href@noop {} {\bibfield  {journal} {\bibinfo  {journal} {Proc. Math. Phys. Eng. Sci.}\ }\textbf {\bibinfo {volume} {348}},\ \bibinfo {pages} {285} (\bibinfo {year} {1976})}\BibitemShut {NoStop}%
\bibitem [{\citenamefont {Gruber}\ \emph {et~al.}(1997)\citenamefont {Gruber}, \citenamefont {Dr{\"a}benstedt}, \citenamefont {Tietz}, \citenamefont {Fleury}, \citenamefont {Wrachtrup},\ and\ \citenamefont {Von~Borczyskowski}}]{gruber1997scanning}%
  \BibitemOpen
  \bibfield  {author} {\bibinfo {author} {\bibfnamefont {A.}~\bibnamefont {Gruber}}, \bibinfo {author} {\bibfnamefont {A.}~\bibnamefont {Dr{\"a}benstedt}}, \bibinfo {author} {\bibfnamefont {C.}~\bibnamefont {Tietz}}, \bibinfo {author} {\bibfnamefont {L.}~\bibnamefont {Fleury}}, \bibinfo {author} {\bibfnamefont {J.}~\bibnamefont {Wrachtrup}},\ and\ \bibinfo {author} {\bibfnamefont {C.}~\bibnamefont {Von~Borczyskowski}},\ }\bibfield  {title} {\bibinfo {title} {Scanning confocal optical microscopy and magnetic resonance on single defect centers},\ }\href@noop {} {\bibfield  {journal} {\bibinfo  {journal} {Science}\ }\textbf {\bibinfo {volume} {276}},\ \bibinfo {pages} {2012} (\bibinfo {year} {1997})}\BibitemShut {NoStop}%
\bibitem [{\citenamefont {Doherty}\ \emph {et~al.}(2013)\citenamefont {Doherty}, \citenamefont {Manson}, \citenamefont {Delaney}, \citenamefont {Jelezko}, \citenamefont {Wrachtrup},\ and\ \citenamefont {Hollenberg}}]{DOHERTY20131}%
  \BibitemOpen
  \bibfield  {author} {\bibinfo {author} {\bibfnamefont {M.~W.}\ \bibnamefont {Doherty}}, \bibinfo {author} {\bibfnamefont {N.~B.}\ \bibnamefont {Manson}}, \bibinfo {author} {\bibfnamefont {P.}~\bibnamefont {Delaney}}, \bibinfo {author} {\bibfnamefont {F.}~\bibnamefont {Jelezko}}, \bibinfo {author} {\bibfnamefont {J.}~\bibnamefont {Wrachtrup}},\ and\ \bibinfo {author} {\bibfnamefont {L.~C.}\ \bibnamefont {Hollenberg}},\ }\bibfield  {title} {\bibinfo {title} {The nitrogen-vacancy colour centre in diamond},\ }\href {https://doi.org/https://doi.org/10.1016/j.physrep.2013.02.001} {\bibfield  {journal} {\bibinfo  {journal} {Phys. Rep.}\ }\textbf {\bibinfo {volume} {528}},\ \bibinfo {pages} {1 } (\bibinfo {year} {2013})},\ \bibinfo {note} {the nitrogen-vacancy colour centre in diamond}\BibitemShut {NoStop}%
\bibitem [{\citenamefont {Wolfowicz}\ \emph {et~al.}(2021)\citenamefont {Wolfowicz}, \citenamefont {Heremans}, \citenamefont {Anderson}, \citenamefont {Kanai}, \citenamefont {Seo}, \citenamefont {Gali}, \citenamefont {Galli},\ and\ \citenamefont {Awschalom}}]{Wolfowicz2021}%
  \BibitemOpen
  \bibfield  {author} {\bibinfo {author} {\bibfnamefont {G.}~\bibnamefont {Wolfowicz}}, \bibinfo {author} {\bibfnamefont {F.~J.}\ \bibnamefont {Heremans}}, \bibinfo {author} {\bibfnamefont {C.~P.}\ \bibnamefont {Anderson}}, \bibinfo {author} {\bibfnamefont {S.}~\bibnamefont {Kanai}}, \bibinfo {author} {\bibfnamefont {H.}~\bibnamefont {Seo}}, \bibinfo {author} {\bibfnamefont {A.}~\bibnamefont {Gali}}, \bibinfo {author} {\bibfnamefont {G.}~\bibnamefont {Galli}},\ and\ \bibinfo {author} {\bibfnamefont {D.~D.}\ \bibnamefont {Awschalom}},\ }\bibfield  {title} {\bibinfo {title} {Quantum guidelines for solid-state spin defects},\ }\href {https://doi.org/10.1038/s41578-021-00306-y} {\bibfield  {journal} {\bibinfo  {journal} {Nature Reviews Materials}\ }\textbf {\bibinfo {volume} {6}},\ \bibinfo {pages} {906} (\bibinfo {year} {2021})}\BibitemShut {NoStop}%
\bibitem [{\citenamefont {Weber}\ \emph {et~al.}(2010)\citenamefont {Weber}, \citenamefont {Koehl}, \citenamefont {Varley}, \citenamefont {Janotti}, \citenamefont {Buckley}, \citenamefont {de~Walle},\ and\ \citenamefont {Awschalom}}]{doi:10.1073/pnas.1003052107}%
  \BibitemOpen
  \bibfield  {author} {\bibinfo {author} {\bibfnamefont {J.~R.}\ \bibnamefont {Weber}}, \bibinfo {author} {\bibfnamefont {W.~F.}\ \bibnamefont {Koehl}}, \bibinfo {author} {\bibfnamefont {J.~B.}\ \bibnamefont {Varley}}, \bibinfo {author} {\bibfnamefont {A.}~\bibnamefont {Janotti}}, \bibinfo {author} {\bibfnamefont {B.~B.}\ \bibnamefont {Buckley}}, \bibinfo {author} {\bibfnamefont {C.~G.~V.}\ \bibnamefont {de~Walle}},\ and\ \bibinfo {author} {\bibfnamefont {D.~D.}\ \bibnamefont {Awschalom}},\ }\bibfield  {title} {\bibinfo {title} {Quantum computing with defects},\ }\href {https://doi.org/10.1073/pnas.1003052107} {\bibfield  {journal} {\bibinfo  {journal} {Proceedings of the National Academy of Sciences}\ }\textbf {\bibinfo {volume} {107}},\ \bibinfo {pages} {8513} (\bibinfo {year} {2010})},\ \Eprint {https://arxiv.org/abs/https://www.pnas.org/doi/pdf/10.1073/pnas.1003052107} {https://www.pnas.org/doi/pdf/10.1073/pnas.1003052107} \BibitemShut {NoStop}%
\bibitem [{\citenamefont {McCullian}\ \emph {et~al.}(2022)\citenamefont {McCullian}, \citenamefont {Cheung}, \citenamefont {Chen},\ and\ \citenamefont {Fuchs}}]{PhysRevApplied.18.064011}%
  \BibitemOpen
  \bibfield  {author} {\bibinfo {author} {\bibfnamefont {B.}~\bibnamefont {McCullian}}, \bibinfo {author} {\bibfnamefont {H.}~\bibnamefont {Cheung}}, \bibinfo {author} {\bibfnamefont {H.}~\bibnamefont {Chen}},\ and\ \bibinfo {author} {\bibfnamefont {G.}~\bibnamefont {Fuchs}},\ }\bibfield  {title} {\bibinfo {title} {Quantifying the spectral diffusion of n-v centers by symmetry},\ }\href {https://doi.org/10.1103/PhysRevApplied.18.064011} {\bibfield  {journal} {\bibinfo  {journal} {Phys. Rev. Appl.}\ }\textbf {\bibinfo {volume} {18}},\ \bibinfo {pages} {064011} (\bibinfo {year} {2022})}\BibitemShut {NoStop}%
\bibitem [{\citenamefont {Alkauskas}\ \emph {et~al.}(2014)\citenamefont {Alkauskas}, \citenamefont {Buckley}, \citenamefont {Awschalom},\ and\ \citenamefont {Van~de Walle}}]{Alkauskas_2014}%
  \BibitemOpen
  \bibfield  {author} {\bibinfo {author} {\bibfnamefont {A.}~\bibnamefont {Alkauskas}}, \bibinfo {author} {\bibfnamefont {B.~B.}\ \bibnamefont {Buckley}}, \bibinfo {author} {\bibfnamefont {D.~D.}\ \bibnamefont {Awschalom}},\ and\ \bibinfo {author} {\bibfnamefont {C.~G.}\ \bibnamefont {Van~de Walle}},\ }\bibfield  {title} {\bibinfo {title} {First-principles theory of the luminescence lineshape for the triplet transition in diamond nv centres},\ }\href {https://doi.org/10.1088/1367-2630/16/7/073026} {\bibfield  {journal} {\bibinfo  {journal} {New Journal of Physics}\ }\textbf {\bibinfo {volume} {16}},\ \bibinfo {pages} {073026} (\bibinfo {year} {2014})}\BibitemShut {NoStop}%
\bibitem [{\citenamefont {DiVincenzo}(2010)}]{DiVincenzo2010}%
  \BibitemOpen
  \bibfield  {author} {\bibinfo {author} {\bibfnamefont {D.}~\bibnamefont {DiVincenzo}},\ }\bibfield  {title} {\bibinfo {title} {Better than excellent},\ }\href {https://doi.org/10.1038/nmat2774} {\bibfield  {journal} {\bibinfo  {journal} {Nature Materials}\ }\textbf {\bibinfo {volume} {9}},\ \bibinfo {pages} {468} (\bibinfo {year} {2010})}\BibitemShut {NoStop}%
\bibitem [{\citenamefont {Zhang}\ \emph {et~al.}(2020)\citenamefont {Zhang}, \citenamefont {Cheng}, \citenamefont {Chou},\ and\ \citenamefont {Gali}}]{10.1063/5.0006075}%
  \BibitemOpen
  \bibfield  {author} {\bibinfo {author} {\bibfnamefont {G.}~\bibnamefont {Zhang}}, \bibinfo {author} {\bibfnamefont {Y.}~\bibnamefont {Cheng}}, \bibinfo {author} {\bibfnamefont {J.-P.}\ \bibnamefont {Chou}},\ and\ \bibinfo {author} {\bibfnamefont {A.}~\bibnamefont {Gali}},\ }\bibfield  {title} {\bibinfo {title} {Material platforms for defect qubits and single-photon emitters},\ }\href {https://doi.org/10.1063/5.0006075} {\bibfield  {journal} {\bibinfo  {journal} {Applied Physics Reviews}\ }\textbf {\bibinfo {volume} {7}},\ \bibinfo {pages} {031308} (\bibinfo {year} {2020})},\ \Eprint {https://arxiv.org/abs/https://pubs.aip.org/aip/apr/article-pdf/doi/10.1063/5.0006075/19742082/031308\_1\_online.pdf} {https://pubs.aip.org/aip/apr/article-pdf/doi/10.1063/5.0006075/19742082/031308\_1\_online.pdf} \BibitemShut {NoStop}%
\bibitem [{\citenamefont {Bassett}\ \emph {et~al.}(2019)\citenamefont {Bassett}, \citenamefont {Alkauskas}, \citenamefont {Exarhos},\ and\ \citenamefont {Fu}}]{BassettAlkauskasExarhosFu}%
  \BibitemOpen
  \bibfield  {author} {\bibinfo {author} {\bibfnamefont {L.~C.}\ \bibnamefont {Bassett}}, \bibinfo {author} {\bibfnamefont {A.}~\bibnamefont {Alkauskas}}, \bibinfo {author} {\bibfnamefont {A.~L.}\ \bibnamefont {Exarhos}},\ and\ \bibinfo {author} {\bibfnamefont {K.-M.~C.}\ \bibnamefont {Fu}},\ }\bibfield  {title} {\bibinfo {title} {Quantum defects by design},\ }\href {https://doi.org/doi:10.1515/nanoph-2019-0211} {\bibfield  {journal} {\bibinfo  {journal} {Nanophotonics}\ }\textbf {\bibinfo {volume} {8}},\ \bibinfo {pages} {1867} (\bibinfo {year} {2019})}\BibitemShut {NoStop}%
\bibitem [{\citenamefont {Ádám Gali}(2019)}]{Gali}%
  \BibitemOpen
  \bibfield  {author} {\bibinfo {author} {\bibnamefont {Ádám Gali}},\ }\bibfield  {title} {\bibinfo {title} {Ab initio theory of the nitrogen-vacancy center in diamond},\ }\href {https://doi.org/doi:10.1515/nanoph-2019-0154} {\bibfield  {journal} {\bibinfo  {journal} {Nanophotonics}\ }\textbf {\bibinfo {volume} {8}},\ \bibinfo {pages} {1907} (\bibinfo {year} {2019})}\BibitemShut {NoStop}%
\bibitem [{fou()}]{fourleaf}%
  \BibitemOpen
  \href {https://news.uga.edu/four-leaf-clover-spotting/} {}\bibinfo {howpublished} {See \url{https://news.uga.edu/four-leaf-clover-spotting/} for information on the commonality of four-leaf clovers and tips for finding them},\ \bibinfo {note} {accessed: 2025-01-15}\BibitemShut {NoStop}%
\bibitem [{\citenamefont {Xiong}\ \emph {et~al.}(2023)\citenamefont {Xiong}, \citenamefont {Bourgois}, \citenamefont {Sheremetyeva}, \citenamefont {Chen}, \citenamefont {Dahliah}, \citenamefont {Song}, \citenamefont {Zheng}, \citenamefont {Griffin}, \citenamefont {Sipahigil},\ and\ \citenamefont {Hautier}}]{doi:10.1126/sciadv.adh8617}%
  \BibitemOpen
  \bibfield  {author} {\bibinfo {author} {\bibfnamefont {Y.}~\bibnamefont {Xiong}}, \bibinfo {author} {\bibfnamefont {C.}~\bibnamefont {Bourgois}}, \bibinfo {author} {\bibfnamefont {N.}~\bibnamefont {Sheremetyeva}}, \bibinfo {author} {\bibfnamefont {W.}~\bibnamefont {Chen}}, \bibinfo {author} {\bibfnamefont {D.}~\bibnamefont {Dahliah}}, \bibinfo {author} {\bibfnamefont {H.}~\bibnamefont {Song}}, \bibinfo {author} {\bibfnamefont {J.}~\bibnamefont {Zheng}}, \bibinfo {author} {\bibfnamefont {S.~M.}\ \bibnamefont {Griffin}}, \bibinfo {author} {\bibfnamefont {A.}~\bibnamefont {Sipahigil}},\ and\ \bibinfo {author} {\bibfnamefont {G.}~\bibnamefont {Hautier}},\ }\bibfield  {title} {\bibinfo {title} {High-throughput identification of spin-photon interfaces in silicon},\ }\href {https://doi.org/10.1126/sciadv.adh8617} {\bibfield  {journal} {\bibinfo  {journal} {Science Advances}\ }\textbf {\bibinfo {volume} {9}},\ \bibinfo {pages} {eadh8617} (\bibinfo {year} {2023})},\ \Eprint
  {https://arxiv.org/abs/https://www.science.org/doi/pdf/10.1126/sciadv.adh8617} {https://www.science.org/doi/pdf/10.1126/sciadv.adh8617} \BibitemShut {NoStop}%
\bibitem [{\citenamefont {Xiong}\ \emph {et~al.}(2024)\citenamefont {Xiong}, \citenamefont {Zheng}, \citenamefont {McBride}, \citenamefont {Zhang}, \citenamefont {Griffin},\ and\ \citenamefont {Hautier}}]{xiong2024discoverytcenterlikequantum}%
  \BibitemOpen
  \bibfield  {author} {\bibinfo {author} {\bibfnamefont {Y.}~\bibnamefont {Xiong}}, \bibinfo {author} {\bibfnamefont {J.}~\bibnamefont {Zheng}}, \bibinfo {author} {\bibfnamefont {S.}~\bibnamefont {McBride}}, \bibinfo {author} {\bibfnamefont {X.}~\bibnamefont {Zhang}}, \bibinfo {author} {\bibfnamefont {S.~M.}\ \bibnamefont {Griffin}},\ and\ \bibinfo {author} {\bibfnamefont {G.}~\bibnamefont {Hautier}},\ }\href {https://arxiv.org/abs/2405.05165} {\bibinfo {title} {Discovery of t center-like quantum defects in silicon}} (\bibinfo {year} {2024}),\ \Eprint {https://arxiv.org/abs/2405.05165} {arXiv:2405.05165 [cond-mat.mtrl-sci]} \BibitemShut {NoStop}%
\bibitem [{\citenamefont {Ivanov}\ \emph {et~al.}(2023)\citenamefont {Ivanov}, \citenamefont {Ivanov}, \citenamefont {Simoni}, \citenamefont {Parajuli}, \citenamefont {Kanté}, \citenamefont {Schenkel},\ and\ \citenamefont {Tan}}]{ivanov2023databasesemiconductorpointdefectproperties}%
  \BibitemOpen
  \bibfield  {author} {\bibinfo {author} {\bibfnamefont {V.}~\bibnamefont {Ivanov}}, \bibinfo {author} {\bibfnamefont {A.}~\bibnamefont {Ivanov}}, \bibinfo {author} {\bibfnamefont {J.}~\bibnamefont {Simoni}}, \bibinfo {author} {\bibfnamefont {P.}~\bibnamefont {Parajuli}}, \bibinfo {author} {\bibfnamefont {B.}~\bibnamefont {Kanté}}, \bibinfo {author} {\bibfnamefont {T.}~\bibnamefont {Schenkel}},\ and\ \bibinfo {author} {\bibfnamefont {L.}~\bibnamefont {Tan}},\ }\href {https://arxiv.org/abs/2303.16283} {\bibinfo {title} {Database of semiconductor point-defect properties for applications in quantum technologies}} (\bibinfo {year} {2023}),\ \Eprint {https://arxiv.org/abs/2303.16283} {arXiv:2303.16283 [quant-ph]} \BibitemShut {NoStop}%
\bibitem [{\citenamefont {Thomas}\ \emph {et~al.}(2024)\citenamefont {Thomas}, \citenamefont {Chen}, \citenamefont {Xiong}, \citenamefont {Barker}, \citenamefont {Zhou}, \citenamefont {Chen}, \citenamefont {Rossi}, \citenamefont {Kelly}, \citenamefont {Yu}, \citenamefont {Zhou}, \citenamefont {Kumari}, \citenamefont {Barnard}, \citenamefont {Robinson}, \citenamefont {Terrones}, \citenamefont {Schwartzberg}, \citenamefont {Ogletree}, \citenamefont {Rotenberg}, \citenamefont {Noack}, \citenamefont {Griffin}, \citenamefont {Raja}, \citenamefont {Strubbe}, \citenamefont {Rignanese}, \citenamefont {Weber-Bargioni},\ and\ \citenamefont {Hautier}}]{Thomas2024}%
  \BibitemOpen
  \bibfield  {author} {\bibinfo {author} {\bibfnamefont {J.~C.}\ \bibnamefont {Thomas}}, \bibinfo {author} {\bibfnamefont {W.}~\bibnamefont {Chen}}, \bibinfo {author} {\bibfnamefont {Y.}~\bibnamefont {Xiong}}, \bibinfo {author} {\bibfnamefont {B.~A.}\ \bibnamefont {Barker}}, \bibinfo {author} {\bibfnamefont {J.}~\bibnamefont {Zhou}}, \bibinfo {author} {\bibfnamefont {W.}~\bibnamefont {Chen}}, \bibinfo {author} {\bibfnamefont {A.}~\bibnamefont {Rossi}}, \bibinfo {author} {\bibfnamefont {N.}~\bibnamefont {Kelly}}, \bibinfo {author} {\bibfnamefont {Z.}~\bibnamefont {Yu}}, \bibinfo {author} {\bibfnamefont {D.}~\bibnamefont {Zhou}}, \bibinfo {author} {\bibfnamefont {S.}~\bibnamefont {Kumari}}, \bibinfo {author} {\bibfnamefont {E.~S.}\ \bibnamefont {Barnard}}, \bibinfo {author} {\bibfnamefont {J.~A.}\ \bibnamefont {Robinson}}, \bibinfo {author} {\bibfnamefont {M.}~\bibnamefont {Terrones}}, \bibinfo {author} {\bibfnamefont {A.}~\bibnamefont {Schwartzberg}}, \bibinfo {author} {\bibfnamefont {D.~F.}\ \bibnamefont
  {Ogletree}}, \bibinfo {author} {\bibfnamefont {E.}~\bibnamefont {Rotenberg}}, \bibinfo {author} {\bibfnamefont {M.~M.}\ \bibnamefont {Noack}}, \bibinfo {author} {\bibfnamefont {S.}~\bibnamefont {Griffin}}, \bibinfo {author} {\bibfnamefont {A.}~\bibnamefont {Raja}}, \bibinfo {author} {\bibfnamefont {D.~A.}\ \bibnamefont {Strubbe}}, \bibinfo {author} {\bibfnamefont {G.-M.}\ \bibnamefont {Rignanese}}, \bibinfo {author} {\bibfnamefont {A.}~\bibnamefont {Weber-Bargioni}},\ and\ \bibinfo {author} {\bibfnamefont {G.}~\bibnamefont {Hautier}},\ }\bibfield  {title} {\bibinfo {title} {A substitutional quantum defect in ws2 discovered by high-throughput computational screening and fabricated by site-selective stm manipulation},\ }\href {https://doi.org/10.1038/s41467-024-47876-3} {\bibfield  {journal} {\bibinfo  {journal} {Nature Communications}\ }\textbf {\bibinfo {volume} {15}},\ \bibinfo {pages} {3556} (\bibinfo {year} {2024})}\BibitemShut {NoStop}%
\bibitem [{\citenamefont {Bertoldo}\ \emph {et~al.}(2022)\citenamefont {Bertoldo}, \citenamefont {Ali}, \citenamefont {Manti},\ and\ \citenamefont {Thygesen}}]{Bertoldo2022}%
  \BibitemOpen
  \bibfield  {author} {\bibinfo {author} {\bibfnamefont {F.}~\bibnamefont {Bertoldo}}, \bibinfo {author} {\bibfnamefont {S.}~\bibnamefont {Ali}}, \bibinfo {author} {\bibfnamefont {S.}~\bibnamefont {Manti}},\ and\ \bibinfo {author} {\bibfnamefont {K.~S.}\ \bibnamefont {Thygesen}},\ }\bibfield  {title} {\bibinfo {title} {Quantum point defects in 2d materials - the qpod database},\ }\href {https://doi.org/10.1038/s41524-022-00730-w} {\bibfield  {journal} {\bibinfo  {journal} {npj Computational Materials}\ }\textbf {\bibinfo {volume} {8}},\ \bibinfo {pages} {56} (\bibinfo {year} {2022})}\BibitemShut {NoStop}%
\bibitem [{\citenamefont {Davidsson}\ \emph {et~al.}(2023)\citenamefont {Davidsson}, \citenamefont {Bertoldo}, \citenamefont {Thygesen},\ and\ \citenamefont {Armiento}}]{abad}%
  \BibitemOpen
  \bibfield  {author} {\bibinfo {author} {\bibfnamefont {J.}~\bibnamefont {Davidsson}}, \bibinfo {author} {\bibfnamefont {F.}~\bibnamefont {Bertoldo}}, \bibinfo {author} {\bibfnamefont {K.~S.}\ \bibnamefont {Thygesen}},\ and\ \bibinfo {author} {\bibfnamefont {R.}~\bibnamefont {Armiento}},\ }\bibfield  {title} {\bibinfo {title} {Absorption versus adsorption: high-throughput computation of impurities in 2d materials},\ }\href {https://doi.org/10.1038/s41699-023-00380-6} {\bibfield  {journal} {\bibinfo  {journal} {npj 2D Materials and Applications}\ }\textbf {\bibinfo {volume} {7}},\ \bibinfo {pages} {26} (\bibinfo {year} {2023})}\BibitemShut {NoStop}%
\bibitem [{\citenamefont {Huang}\ \emph {et~al.}(2023)\citenamefont {Huang}, \citenamefont {Lukin}, \citenamefont {Faleev}, \citenamefont {Kazeev}, \citenamefont {Al-Maeeni}, \citenamefont {Andreeva}, \citenamefont {Ustyuzhanin}, \citenamefont {Tormasov}, \citenamefont {Castro~Neto},\ and\ \citenamefont {Novoselov}}]{Huang2023}%
  \BibitemOpen
  \bibfield  {author} {\bibinfo {author} {\bibfnamefont {P.}~\bibnamefont {Huang}}, \bibinfo {author} {\bibfnamefont {R.}~\bibnamefont {Lukin}}, \bibinfo {author} {\bibfnamefont {M.}~\bibnamefont {Faleev}}, \bibinfo {author} {\bibfnamefont {N.}~\bibnamefont {Kazeev}}, \bibinfo {author} {\bibfnamefont {A.~R.}\ \bibnamefont {Al-Maeeni}}, \bibinfo {author} {\bibfnamefont {D.~V.}\ \bibnamefont {Andreeva}}, \bibinfo {author} {\bibfnamefont {A.}~\bibnamefont {Ustyuzhanin}}, \bibinfo {author} {\bibfnamefont {A.}~\bibnamefont {Tormasov}}, \bibinfo {author} {\bibfnamefont {A.~H.}\ \bibnamefont {Castro~Neto}},\ and\ \bibinfo {author} {\bibfnamefont {K.~S.}\ \bibnamefont {Novoselov}},\ }\bibfield  {title} {\bibinfo {title} {Unveiling the complex structure-property correlation of defects in 2d materials based on high throughput datasets},\ }\href {https://doi.org/10.1038/s41699-023-00369-1} {\bibfield  {journal} {\bibinfo  {journal} {npj 2D Materials and Applications}\ }\textbf {\bibinfo {volume} {7}},\ \bibinfo {pages}
  {6} (\bibinfo {year} {2023})}\BibitemShut {NoStop}%
\bibitem [{\citenamefont {Iv\'ady}\ \emph {et~al.}(2017)\citenamefont {Iv\'ady}, \citenamefont {Davidsson}, \citenamefont {Son}, \citenamefont {Ohshima}, \citenamefont {Abrikosov},\ and\ \citenamefont {Gali}}]{vacancypaper}%
  \BibitemOpen
  \bibfield  {author} {\bibinfo {author} {\bibfnamefont {V.}~\bibnamefont {Iv\'ady}}, \bibinfo {author} {\bibfnamefont {J.}~\bibnamefont {Davidsson}}, \bibinfo {author} {\bibfnamefont {N.~T.}\ \bibnamefont {Son}}, \bibinfo {author} {\bibfnamefont {T.}~\bibnamefont {Ohshima}}, \bibinfo {author} {\bibfnamefont {I.~A.}\ \bibnamefont {Abrikosov}},\ and\ \bibinfo {author} {\bibfnamefont {A.}~\bibnamefont {Gali}},\ }\bibfield  {title} {\bibinfo {title} {Identification of si-vacancy related room-temperature qubits in $4h$ silicon carbide},\ }\href {https://doi.org/10.1103/PhysRevB.96.161114} {\bibfield  {journal} {\bibinfo  {journal} {Phys. Rev. B}\ }\textbf {\bibinfo {volume} {96}},\ \bibinfo {pages} {161114} (\bibinfo {year} {2017})}\BibitemShut {NoStop}%
\bibitem [{\citenamefont {Davidsson}\ \emph {et~al.}(2018)\citenamefont {Davidsson}, \citenamefont {Iv{\'{a}}dy}, \citenamefont {Armiento}, \citenamefont {Son}, \citenamefont {Gali},\ and\ \citenamefont {Abrikosov}}]{methodologypaper}%
  \BibitemOpen
  \bibfield  {author} {\bibinfo {author} {\bibfnamefont {J.}~\bibnamefont {Davidsson}}, \bibinfo {author} {\bibfnamefont {V.}~\bibnamefont {Iv{\'{a}}dy}}, \bibinfo {author} {\bibfnamefont {R.}~\bibnamefont {Armiento}}, \bibinfo {author} {\bibfnamefont {N.~T.}\ \bibnamefont {Son}}, \bibinfo {author} {\bibfnamefont {A.}~\bibnamefont {Gali}},\ and\ \bibinfo {author} {\bibfnamefont {I.~A.}\ \bibnamefont {Abrikosov}},\ }\bibfield  {title} {\bibinfo {title} {First principles predictions of magneto-optical data for semiconductor point defect identification: the case of divacancy defects in 4h{\textendash}{SiC}},\ }\href {https://doi.org/10.1088/1367-2630/aaa752} {\bibfield  {journal} {\bibinfo  {journal} {New Journal of Physics}\ }\textbf {\bibinfo {volume} {20}},\ \bibinfo {pages} {023035} (\bibinfo {year} {2018})}\BibitemShut {NoStop}%
\bibitem [{\citenamefont {Davidsson}\ \emph {et~al.}(2019)\citenamefont {Davidsson}, \citenamefont {Ivády}, \citenamefont {Armiento}, \citenamefont {Ohshima}, \citenamefont {Son}, \citenamefont {Gali},\ and\ \citenamefont {Abrikosov}}]{6Hpaper}%
  \BibitemOpen
  \bibfield  {author} {\bibinfo {author} {\bibfnamefont {J.}~\bibnamefont {Davidsson}}, \bibinfo {author} {\bibfnamefont {V.}~\bibnamefont {Ivády}}, \bibinfo {author} {\bibfnamefont {R.}~\bibnamefont {Armiento}}, \bibinfo {author} {\bibfnamefont {T.}~\bibnamefont {Ohshima}}, \bibinfo {author} {\bibfnamefont {N.~T.}\ \bibnamefont {Son}}, \bibinfo {author} {\bibfnamefont {A.}~\bibnamefont {Gali}},\ and\ \bibinfo {author} {\bibfnamefont {I.~A.}\ \bibnamefont {Abrikosov}},\ }\bibfield  {title} {\bibinfo {title} {Identification of divacancy and silicon vacancy qubits in 6h-sic},\ }\href {https://doi.org/10.1063/1.5083031} {\bibfield  {journal} {\bibinfo  {journal} {Applied Physics Letters}\ }\textbf {\bibinfo {volume} {114}},\ \bibinfo {pages} {112107} (\bibinfo {year} {2019})},\ \Eprint {https://arxiv.org/abs/https://doi.org/10.1063/1.5083031} {https://doi.org/10.1063/1.5083031} \BibitemShut {NoStop}%
\bibitem [{\citenamefont {Davidsson}(2020)}]{mypaper}%
  \BibitemOpen
  \bibfield  {author} {\bibinfo {author} {\bibfnamefont {J.}~\bibnamefont {Davidsson}},\ }\bibfield  {title} {\bibinfo {title} {Theoretical polarization of zero phonon lines in point defects},\ }\href {https://doi.org/10.1088/1361-648x/ab94f4} {\bibfield  {journal} {\bibinfo  {journal} {Journal of Physics: Condensed Matter}\ }\textbf {\bibinfo {volume} {32}},\ \bibinfo {pages} {385502} (\bibinfo {year} {2020})}\BibitemShut {NoStop}%
\bibitem [{\citenamefont {Davidsson}(2021)}]{thesis}%
  \BibitemOpen
  \bibfield  {author} {\bibinfo {author} {\bibfnamefont {J.}~\bibnamefont {Davidsson}},\ }\emph {\bibinfo {title} {Color Centers in Semiconductors for Quantum Applications: A High-Throughput Search of Point Defects in SiC}},\ \href {https://doi.org/10.3384/diss.diva-173108} {Ph.D. thesis},\ \bibinfo  {school} {Link{\"o}ping University Electronic Press} (\bibinfo {year} {2021})\BibitemShut {NoStop}%
\bibitem [{\citenamefont {Davidsson}\ \emph {et~al.}(2021)\citenamefont {Davidsson}, \citenamefont {Ivády}, \citenamefont {Armiento},\ and\ \citenamefont {Abrikosov}}]{ADAQpaper}%
  \BibitemOpen
  \bibfield  {author} {\bibinfo {author} {\bibfnamefont {J.}~\bibnamefont {Davidsson}}, \bibinfo {author} {\bibfnamefont {V.}~\bibnamefont {Ivády}}, \bibinfo {author} {\bibfnamefont {R.}~\bibnamefont {Armiento}},\ and\ \bibinfo {author} {\bibfnamefont {I.~A.}\ \bibnamefont {Abrikosov}},\ }\bibfield  {title} {\bibinfo {title} {Adaq: Automatic workflows for magneto-optical properties of point defects in semiconductors},\ }\href {https://doi.org/https://doi.org/10.1016/j.cpc.2021.108091} {\bibfield  {journal} {\bibinfo  {journal} {Computer Physics Communications}\ }\textbf {\bibinfo {volume} {269}},\ \bibinfo {pages} {108091} (\bibinfo {year} {2021})}\BibitemShut {NoStop}%
\bibitem [{\citenamefont {Armiento}(2020)}]{armientoDatabaseDrivenHighThroughputCalculations2020}%
  \BibitemOpen
  \bibfield  {author} {\bibinfo {author} {\bibfnamefont {R.}~\bibnamefont {Armiento}},\ }\bibfield  {title} {\bibinfo {title} {Database-{{Driven High-Throughput Calculations}} and {{Machine Learning Models}} for {{Materials Design}}},\ }in\ \href@noop {} {\emph {\bibinfo {booktitle} {Machine {{Learning Meets Quantum Physics}}}}},\ \bibinfo {series} {Lecture {{Notes}} in {{Physics}}}, Vol.\ \bibinfo {volume} {968},\ \bibinfo {editor} {edited by\ \bibinfo {editor} {\bibfnamefont {K.~T.}\ \bibnamefont {Sch{\"u}tt}}, \bibinfo {editor} {\bibfnamefont {S.}~\bibnamefont {Chmiela}}, \bibinfo {editor} {\bibfnamefont {O.~A.}\ \bibnamefont {{von Lilienfeld}}}, \bibinfo {editor} {\bibfnamefont {A.}~\bibnamefont {Tkatchenko}}, \bibinfo {editor} {\bibfnamefont {K.}~\bibnamefont {Tsuda}},\ and\ \bibinfo {editor} {\bibfnamefont {K.-R.}\ \bibnamefont {M{\"u}ller}}}\ (\bibinfo  {publisher} {{Springer International Publishing}},\ \bibinfo {address} {{Cham}},\ \bibinfo {year} {2020})\BibitemShut {NoStop}%
\bibitem [{\citenamefont {Kresse}\ and\ \citenamefont {Hafner}(1994)}]{VASP}%
  \BibitemOpen
  \bibfield  {author} {\bibinfo {author} {\bibfnamefont {G.}~\bibnamefont {Kresse}}\ and\ \bibinfo {author} {\bibfnamefont {J.}~\bibnamefont {Hafner}},\ }\bibfield  {title} {\bibinfo {title} {\textit{Ab initio} molecular-dynamics simulation of the liquid-metal–amorphous-semiconductor transition in germanium},\ }\href {https://doi.org/10.1103/PhysRevB.49.14251} {\bibfield  {journal} {\bibinfo  {journal} {Phys. Rev. B}\ }\textbf {\bibinfo {volume} {49}},\ \bibinfo {pages} {14251} (\bibinfo {year} {1994})}\BibitemShut {NoStop}%
\bibitem [{\citenamefont {Kresse}\ and\ \citenamefont {Furthm\"uller}(1996)}]{VASP2}%
  \BibitemOpen
  \bibfield  {author} {\bibinfo {author} {\bibfnamefont {G.}~\bibnamefont {Kresse}}\ and\ \bibinfo {author} {\bibfnamefont {J.}~\bibnamefont {Furthm\"uller}},\ }\bibfield  {title} {\bibinfo {title} {Efficient iterative schemes for \textit{ab initio} total-energy calculations using a plane-wave basis set},\ }\href {https://doi.org/10.1103/PhysRevB.54.11169} {\bibfield  {journal} {\bibinfo  {journal} {Phys. Rev. B}\ }\textbf {\bibinfo {volume} {54}},\ \bibinfo {pages} {11169} (\bibinfo {year} {1996})}\BibitemShut {NoStop}%
\bibitem [{\citenamefont {Bl\"ochl}(1994)}]{PAW}%
  \BibitemOpen
  \bibfield  {author} {\bibinfo {author} {\bibfnamefont {P.~E.}\ \bibnamefont {Bl\"ochl}},\ }\bibfield  {title} {\bibinfo {title} {Projector augmented-wave method},\ }\href {https://doi.org/10.1103/PhysRevB.50.17953} {\bibfield  {journal} {\bibinfo  {journal} {Phys. Rev. B}\ }\textbf {\bibinfo {volume} {50}},\ \bibinfo {pages} {17953} (\bibinfo {year} {1994})}\BibitemShut {NoStop}%
\bibitem [{\citenamefont {Kresse}\ and\ \citenamefont {Joubert}(1999)}]{Kresse99}%
  \BibitemOpen
  \bibfield  {author} {\bibinfo {author} {\bibfnamefont {G.}~\bibnamefont {Kresse}}\ and\ \bibinfo {author} {\bibfnamefont {D.}~\bibnamefont {Joubert}},\ }\bibfield  {title} {\bibinfo {title} {From ultrasoft pseudopotentials to the projector augmented-wave method},\ }\href {https://doi.org/10.1103/PhysRevB.59.1758} {\bibfield  {journal} {\bibinfo  {journal} {Phys. Rev. B}\ }\textbf {\bibinfo {volume} {59}},\ \bibinfo {pages} {1758} (\bibinfo {year} {1999})}\BibitemShut {NoStop}%
\bibitem [{\citenamefont {Perdew}\ \emph {et~al.}(1996)\citenamefont {Perdew}, \citenamefont {Burke},\ and\ \citenamefont {Ernzerhof}}]{perdew1996generalized}%
  \BibitemOpen
  \bibfield  {author} {\bibinfo {author} {\bibfnamefont {J.~P.}\ \bibnamefont {Perdew}}, \bibinfo {author} {\bibfnamefont {K.}~\bibnamefont {Burke}},\ and\ \bibinfo {author} {\bibfnamefont {M.}~\bibnamefont {Ernzerhof}},\ }\bibfield  {title} {\bibinfo {title} {Generalized gradient approximation made simple},\ }\href@noop {} {\bibfield  {journal} {\bibinfo  {journal} {Physical review letters}\ }\textbf {\bibinfo {volume} {77}},\ \bibinfo {pages} {3865} (\bibinfo {year} {1996})}\BibitemShut {NoStop}%
\bibitem [{\citenamefont {Kramer}\ and\ \citenamefont {MacKinnon}(1993)}]{IPR1}%
  \BibitemOpen
  \bibfield  {author} {\bibinfo {author} {\bibfnamefont {B.}~\bibnamefont {Kramer}}\ and\ \bibinfo {author} {\bibfnamefont {A.}~\bibnamefont {MacKinnon}},\ }\bibfield  {title} {\bibinfo {title} {Localization: theory and experiment},\ }\href {https://doi.org/10.1088/0034-4885/56/12/001} {\bibfield  {journal} {\bibinfo  {journal} {Rep. Prog. Phys.}\ }\textbf {\bibinfo {volume} {56}},\ \bibinfo {pages} {1469} (\bibinfo {year} {1993})}\BibitemShut {NoStop}%
\bibitem [{\citenamefont {Kaduk}\ \emph {et~al.}(2012)\citenamefont {Kaduk}, \citenamefont {Kowalczyk},\ and\ \citenamefont {Van~Voorhis}}]{cDFT}%
  \BibitemOpen
  \bibfield  {author} {\bibinfo {author} {\bibfnamefont {B.}~\bibnamefont {Kaduk}}, \bibinfo {author} {\bibfnamefont {T.}~\bibnamefont {Kowalczyk}},\ and\ \bibinfo {author} {\bibfnamefont {T.}~\bibnamefont {Van~Voorhis}},\ }\bibfield  {title} {\bibinfo {title} {Constrained density functional theory},\ }\href {https://doi.org/10.1021/cr200148b} {\bibfield  {journal} {\bibinfo  {journal} {Chem. Rev.}\ }\textbf {\bibinfo {volume} {112}},\ \bibinfo {pages} {321} (\bibinfo {year} {2012})},\ \bibinfo {note} {pMID: 22077560},\ \Eprint {https://arxiv.org/abs/https://doi.org/10.1021/cr200148b} {https://doi.org/10.1021/cr200148b} \BibitemShut {NoStop}%
\bibitem [{\citenamefont {Evans}\ \emph {et~al.}(2024)\citenamefont {Evans}, \citenamefont {Bergsma}, \citenamefont {Merkys}, \citenamefont {Andersen}, \citenamefont {Andersson}, \citenamefont {Beltrán}, \citenamefont {Blokhin}, \citenamefont {Boland}, \citenamefont {Castañeda~Balderas}, \citenamefont {Choudhary}, \citenamefont {Díaz~Díaz}, \citenamefont {Domínguez~García}, \citenamefont {Eckert}, \citenamefont {Eimre}, \citenamefont {Fuentes~Montero}, \citenamefont {Krajewski}, \citenamefont {Mortensen}, \citenamefont {Nápoles~Duarte}, \citenamefont {Pietryga}, \citenamefont {Qi}, \citenamefont {Trejo~Carrillo}, \citenamefont {Vaitkus}, \citenamefont {Yu}, \citenamefont {Zettel}, \citenamefont {de~Castro}, \citenamefont {Carlsson}, \citenamefont {Cerqueira}, \citenamefont {Divilov}, \citenamefont {Hajiyani}, \citenamefont {Hanke}, \citenamefont {Jose}, \citenamefont {Oses}, \citenamefont {Riebesell}, \citenamefont {Schmidt}, \citenamefont {Winston}, \citenamefont {Xie}, \citenamefont {Yang},
  \citenamefont {Bonella}, \citenamefont {Botti}, \citenamefont {Curtarolo}, \citenamefont {Draxl}, \citenamefont {Fuentes~Cobas}, \citenamefont {Hospital}, \citenamefont {Liu}, \citenamefont {Marques}, \citenamefont {Marzari}, \citenamefont {Morris}, \citenamefont {Ong}, \citenamefont {Orozco}, \citenamefont {Persson}, \citenamefont {Thygesen}, \citenamefont {Wolverton}, \citenamefont {Scheidgen}, \citenamefont {Toher}, \citenamefont {Conduit}, \citenamefont {Pizzi}, \citenamefont {Gražulis}, \citenamefont {Rignanese},\ and\ \citenamefont {Armiento}}]{D4DD00039K}%
  \BibitemOpen
  \bibfield  {author} {\bibinfo {author} {\bibfnamefont {M.~L.}\ \bibnamefont {Evans}}, \bibinfo {author} {\bibfnamefont {J.}~\bibnamefont {Bergsma}}, \bibinfo {author} {\bibfnamefont {A.}~\bibnamefont {Merkys}}, \bibinfo {author} {\bibfnamefont {C.~W.}\ \bibnamefont {Andersen}}, \bibinfo {author} {\bibfnamefont {O.~B.}\ \bibnamefont {Andersson}}, \bibinfo {author} {\bibfnamefont {D.}~\bibnamefont {Beltrán}}, \bibinfo {author} {\bibfnamefont {E.}~\bibnamefont {Blokhin}}, \bibinfo {author} {\bibfnamefont {T.~M.}\ \bibnamefont {Boland}}, \bibinfo {author} {\bibfnamefont {R.}~\bibnamefont {Castañeda~Balderas}}, \bibinfo {author} {\bibfnamefont {K.}~\bibnamefont {Choudhary}}, \bibinfo {author} {\bibfnamefont {A.}~\bibnamefont {Díaz~Díaz}}, \bibinfo {author} {\bibfnamefont {R.}~\bibnamefont {Domínguez~García}}, \bibinfo {author} {\bibfnamefont {H.}~\bibnamefont {Eckert}}, \bibinfo {author} {\bibfnamefont {K.}~\bibnamefont {Eimre}}, \bibinfo {author} {\bibfnamefont {M.~E.}\ \bibnamefont {Fuentes~Montero}},
  \bibinfo {author} {\bibfnamefont {A.~M.}\ \bibnamefont {Krajewski}}, \bibinfo {author} {\bibfnamefont {J.~J.}\ \bibnamefont {Mortensen}}, \bibinfo {author} {\bibfnamefont {J.~M.}\ \bibnamefont {Nápoles~Duarte}}, \bibinfo {author} {\bibfnamefont {J.}~\bibnamefont {Pietryga}}, \bibinfo {author} {\bibfnamefont {J.}~\bibnamefont {Qi}}, \bibinfo {author} {\bibfnamefont {F.~d.~J.}\ \bibnamefont {Trejo~Carrillo}}, \bibinfo {author} {\bibfnamefont {A.}~\bibnamefont {Vaitkus}}, \bibinfo {author} {\bibfnamefont {J.}~\bibnamefont {Yu}}, \bibinfo {author} {\bibfnamefont {A.}~\bibnamefont {Zettel}}, \bibinfo {author} {\bibfnamefont {P.~B.}\ \bibnamefont {de~Castro}}, \bibinfo {author} {\bibfnamefont {J.}~\bibnamefont {Carlsson}}, \bibinfo {author} {\bibfnamefont {T.~F.~T.}\ \bibnamefont {Cerqueira}}, \bibinfo {author} {\bibfnamefont {S.}~\bibnamefont {Divilov}}, \bibinfo {author} {\bibfnamefont {H.}~\bibnamefont {Hajiyani}}, \bibinfo {author} {\bibfnamefont {F.}~\bibnamefont {Hanke}}, \bibinfo {author} {\bibfnamefont
  {K.}~\bibnamefont {Jose}}, \bibinfo {author} {\bibfnamefont {C.}~\bibnamefont {Oses}}, \bibinfo {author} {\bibfnamefont {J.}~\bibnamefont {Riebesell}}, \bibinfo {author} {\bibfnamefont {J.}~\bibnamefont {Schmidt}}, \bibinfo {author} {\bibfnamefont {D.}~\bibnamefont {Winston}}, \bibinfo {author} {\bibfnamefont {C.}~\bibnamefont {Xie}}, \bibinfo {author} {\bibfnamefont {X.}~\bibnamefont {Yang}}, \bibinfo {author} {\bibfnamefont {S.}~\bibnamefont {Bonella}}, \bibinfo {author} {\bibfnamefont {S.}~\bibnamefont {Botti}}, \bibinfo {author} {\bibfnamefont {S.}~\bibnamefont {Curtarolo}}, \bibinfo {author} {\bibfnamefont {C.}~\bibnamefont {Draxl}}, \bibinfo {author} {\bibfnamefont {L.~E.}\ \bibnamefont {Fuentes~Cobas}}, \bibinfo {author} {\bibfnamefont {A.}~\bibnamefont {Hospital}}, \bibinfo {author} {\bibfnamefont {Z.-K.}\ \bibnamefont {Liu}}, \bibinfo {author} {\bibfnamefont {M.~A.~L.}\ \bibnamefont {Marques}}, \bibinfo {author} {\bibfnamefont {N.}~\bibnamefont {Marzari}}, \bibinfo {author} {\bibfnamefont {A.~J.}\
  \bibnamefont {Morris}}, \bibinfo {author} {\bibfnamefont {S.~P.}\ \bibnamefont {Ong}}, \bibinfo {author} {\bibfnamefont {M.}~\bibnamefont {Orozco}}, \bibinfo {author} {\bibfnamefont {K.~A.}\ \bibnamefont {Persson}}, \bibinfo {author} {\bibfnamefont {K.~S.}\ \bibnamefont {Thygesen}}, \bibinfo {author} {\bibfnamefont {C.}~\bibnamefont {Wolverton}}, \bibinfo {author} {\bibfnamefont {M.}~\bibnamefont {Scheidgen}}, \bibinfo {author} {\bibfnamefont {C.}~\bibnamefont {Toher}}, \bibinfo {author} {\bibfnamefont {G.~J.}\ \bibnamefont {Conduit}}, \bibinfo {author} {\bibfnamefont {G.}~\bibnamefont {Pizzi}}, \bibinfo {author} {\bibfnamefont {S.}~\bibnamefont {Gražulis}}, \bibinfo {author} {\bibfnamefont {G.-M.}\ \bibnamefont {Rignanese}},\ and\ \bibinfo {author} {\bibfnamefont {R.}~\bibnamefont {Armiento}},\ }\bibfield  {title} {\bibinfo {title} {Developments and applications of the optimade api for materials discovery{,} design{,} and data exchange},\ }\href {https://doi.org/10.1039/D4DD00039K} {\bibfield  {journal}
  {\bibinfo  {journal} {Digital Discovery}\ }\textbf {\bibinfo {volume} {3}},\ \bibinfo {pages} {1509} (\bibinfo {year} {2024})}\BibitemShut {NoStop}%
\bibitem [{\citenamefont {Davidsson}\ \emph {et~al.}(2022)\citenamefont {Davidsson}, \citenamefont {Babar}, \citenamefont {Shafizadeh}, \citenamefont {Ivanov}, \citenamefont {Ivády}, \citenamefont {Armiento},\ and\ \citenamefont {Abrikosov}}]{modvac}%
  \BibitemOpen
  \bibfield  {author} {\bibinfo {author} {\bibfnamefont {J.}~\bibnamefont {Davidsson}}, \bibinfo {author} {\bibfnamefont {R.}~\bibnamefont {Babar}}, \bibinfo {author} {\bibfnamefont {D.}~\bibnamefont {Shafizadeh}}, \bibinfo {author} {\bibfnamefont {I.~G.}\ \bibnamefont {Ivanov}}, \bibinfo {author} {\bibfnamefont {V.}~\bibnamefont {Ivády}}, \bibinfo {author} {\bibfnamefont {R.}~\bibnamefont {Armiento}},\ and\ \bibinfo {author} {\bibfnamefont {I.~A.}\ \bibnamefont {Abrikosov}},\ }\bibfield  {title} {\bibinfo {title} {Exhaustive characterization of modified si vacancies in 4h-sic},\ }\href {https://doi.org/doi:10.1515/nanoph-2022-0400} {\bibfield  {journal} {\bibinfo  {journal} {Nanophotonics}\ }\textbf {\bibinfo {volume} {11}},\ \bibinfo {pages} {4565} (\bibinfo {year} {2022})}\BibitemShut {NoStop}%
\bibitem [{\citenamefont {Bradac}\ \emph {et~al.}(2019)\citenamefont {Bradac}, \citenamefont {Gao}, \citenamefont {Forneris}, \citenamefont {Trusheim},\ and\ \citenamefont {Aharonovich}}]{Bradac2019}%
  \BibitemOpen
  \bibfield  {author} {\bibinfo {author} {\bibfnamefont {C.}~\bibnamefont {Bradac}}, \bibinfo {author} {\bibfnamefont {W.}~\bibnamefont {Gao}}, \bibinfo {author} {\bibfnamefont {J.}~\bibnamefont {Forneris}}, \bibinfo {author} {\bibfnamefont {M.~E.}\ \bibnamefont {Trusheim}},\ and\ \bibinfo {author} {\bibfnamefont {I.}~\bibnamefont {Aharonovich}},\ }\bibfield  {title} {\bibinfo {title} {Quantum nanophotonics with group iv defects in diamond},\ }\href {https://doi.org/10.1038/s41467-019-13332-w} {\bibfield  {journal} {\bibinfo  {journal} {Nature Communications}\ }\textbf {\bibinfo {volume} {10}},\ \bibinfo {pages} {5625} (\bibinfo {year} {2019})}\BibitemShut {NoStop}%
\bibitem [{\citenamefont {Davidsson}\ \emph {et~al.}(2024{\natexlab{a}})\citenamefont {Davidsson}, \citenamefont {Stenlund}, \citenamefont {Parackal}, \citenamefont {Armiento},\ and\ \citenamefont {Abrikosov}}]{davidsson2023na}%
  \BibitemOpen
  \bibfield  {author} {\bibinfo {author} {\bibfnamefont {J.}~\bibnamefont {Davidsson}}, \bibinfo {author} {\bibfnamefont {W.}~\bibnamefont {Stenlund}}, \bibinfo {author} {\bibfnamefont {A.~S.}\ \bibnamefont {Parackal}}, \bibinfo {author} {\bibfnamefont {R.}~\bibnamefont {Armiento}},\ and\ \bibinfo {author} {\bibfnamefont {I.~A.}\ \bibnamefont {Abrikosov}},\ }\bibfield  {title} {\bibinfo {title} {Na in diamond: high spin defects revealed by the adaq high-throughput computational database},\ }\href {https://doi.org/10.1038/s41524-024-01292-9} {\bibfield  {journal} {\bibinfo  {journal} {npj Computational Materials}\ }\textbf {\bibinfo {volume} {10}},\ \bibinfo {pages} {109} (\bibinfo {year} {2024}{\natexlab{a}})}\BibitemShut {NoStop}%
\bibitem [{\citenamefont {Thiering}\ and\ \citenamefont {Gali}(2018)}]{PhysRevX.8.021063}%
  \BibitemOpen
  \bibfield  {author} {\bibinfo {author} {\bibfnamefont {G.~m.~H.}\ \bibnamefont {Thiering}}\ and\ \bibinfo {author} {\bibfnamefont {A.}~\bibnamefont {Gali}},\ }\bibfield  {title} {\bibinfo {title} {Ab initio magneto-optical spectrum of group-iv vacancy color centers in diamond},\ }\href {https://doi.org/10.1103/PhysRevX.8.021063} {\bibfield  {journal} {\bibinfo  {journal} {Phys. Rev. X}\ }\textbf {\bibinfo {volume} {8}},\ \bibinfo {pages} {021063} (\bibinfo {year} {2018})}\BibitemShut {NoStop}%
\bibitem [{\citenamefont {Falk}\ \emph {et~al.}(2013)\citenamefont {Falk}, \citenamefont {Buckley}, \citenamefont {Calusine}, \citenamefont {Koehl}, \citenamefont {Dobrovitski}, \citenamefont {Politi}, \citenamefont {Zorman}, \citenamefont {Feng},\ and\ \citenamefont {Awschalom}}]{Falk2013}%
  \BibitemOpen
  \bibfield  {author} {\bibinfo {author} {\bibfnamefont {A.~L.}\ \bibnamefont {Falk}}, \bibinfo {author} {\bibfnamefont {B.~B.}\ \bibnamefont {Buckley}}, \bibinfo {author} {\bibfnamefont {G.}~\bibnamefont {Calusine}}, \bibinfo {author} {\bibfnamefont {W.~F.}\ \bibnamefont {Koehl}}, \bibinfo {author} {\bibfnamefont {V.~V.}\ \bibnamefont {Dobrovitski}}, \bibinfo {author} {\bibfnamefont {A.}~\bibnamefont {Politi}}, \bibinfo {author} {\bibfnamefont {C.~A.}\ \bibnamefont {Zorman}}, \bibinfo {author} {\bibfnamefont {P.~X.-L.}\ \bibnamefont {Feng}},\ and\ \bibinfo {author} {\bibfnamefont {D.~D.}\ \bibnamefont {Awschalom}},\ }\bibfield  {title} {\bibinfo {title} {Polytype control of spin qubits in silicon carbide},\ }\href {https://doi.org/10.1038/ncomms2854} {\bibfield  {journal} {\bibinfo  {journal} {Nature Communications}\ }\textbf {\bibinfo {volume} {4}},\ \bibinfo {pages} {1819} (\bibinfo {year} {2013})}\BibitemShut {NoStop}%
\bibitem [{\citenamefont {Wagner}\ \emph {et~al.}(2000)\citenamefont {Wagner}, \citenamefont {Magnusson}, \citenamefont {Chen}, \citenamefont {Janz\'en}, \citenamefont {S\"orman}, \citenamefont {Hallin},\ and\ \citenamefont {Lindstr\"om}}]{PhysRevB.62.16555}%
  \BibitemOpen
  \bibfield  {author} {\bibinfo {author} {\bibfnamefont {M.}~\bibnamefont {Wagner}}, \bibinfo {author} {\bibfnamefont {B.}~\bibnamefont {Magnusson}}, \bibinfo {author} {\bibfnamefont {W.~M.}\ \bibnamefont {Chen}}, \bibinfo {author} {\bibfnamefont {E.}~\bibnamefont {Janz\'en}}, \bibinfo {author} {\bibfnamefont {E.}~\bibnamefont {S\"orman}}, \bibinfo {author} {\bibfnamefont {C.}~\bibnamefont {Hallin}},\ and\ \bibinfo {author} {\bibfnamefont {J.~L.}\ \bibnamefont {Lindstr\"om}},\ }\bibfield  {title} {\bibinfo {title} {Electronic structure of the neutral silicon vacancy in $4h$ and $6h$ sic},\ }\href {https://doi.org/10.1103/PhysRevB.62.16555} {\bibfield  {journal} {\bibinfo  {journal} {Phys. Rev. B}\ }\textbf {\bibinfo {volume} {62}},\ \bibinfo {pages} {16555} (\bibinfo {year} {2000})}\BibitemShut {NoStop}%
\bibitem [{\citenamefont {Bulancea-Lindvall}\ \emph {et~al.}(2024)\citenamefont {Bulancea-Lindvall}, \citenamefont {Davidsson}, \citenamefont {Ivanov}, \citenamefont {Gali}, \citenamefont {Iv\'ady}, \citenamefont {Armiento},\ and\ \citenamefont {Abrikosov}}]{cavpaper}%
  \BibitemOpen
  \bibfield  {author} {\bibinfo {author} {\bibfnamefont {O.}~\bibnamefont {Bulancea-Lindvall}}, \bibinfo {author} {\bibfnamefont {J.}~\bibnamefont {Davidsson}}, \bibinfo {author} {\bibfnamefont {I.~G.}\ \bibnamefont {Ivanov}}, \bibinfo {author} {\bibfnamefont {A.}~\bibnamefont {Gali}}, \bibinfo {author} {\bibfnamefont {V.}~\bibnamefont {Iv\'ady}}, \bibinfo {author} {\bibfnamefont {R.}~\bibnamefont {Armiento}},\ and\ \bibinfo {author} {\bibfnamefont {I.~A.}\ \bibnamefont {Abrikosov}},\ }\bibfield  {title} {\bibinfo {title} {Temperature dependence of the ab lines and optical properties of the carbon--antisite-vacancy pair in 4h-sic},\ }\href {https://doi.org/10.1103/PhysRevApplied.22.034056} {\bibfield  {journal} {\bibinfo  {journal} {Phys. Rev. Appl.}\ }\textbf {\bibinfo {volume} {22}},\ \bibinfo {pages} {034056} (\bibinfo {year} {2024})}\BibitemShut {NoStop}%
\bibitem [{Note1()}]{Note1}%
  \BibitemOpen
  \bibinfo {note} {Originally there were 21,607 defects, but a bug created duplicates. After removing these duplicates, 8,450 unique defects remained.}\BibitemShut {Stop}%
\bibitem [{\citenamefont {Bulancea-Lindvall}\ \emph {et~al.}(2023)\citenamefont {Bulancea-Lindvall}, \citenamefont {Davidsson}, \citenamefont {Armiento},\ and\ \citenamefont {Abrikosov}}]{clvpaper}%
  \BibitemOpen
  \bibfield  {author} {\bibinfo {author} {\bibfnamefont {O.}~\bibnamefont {Bulancea-Lindvall}}, \bibinfo {author} {\bibfnamefont {J.}~\bibnamefont {Davidsson}}, \bibinfo {author} {\bibfnamefont {R.}~\bibnamefont {Armiento}},\ and\ \bibinfo {author} {\bibfnamefont {I.~A.}\ \bibnamefont {Abrikosov}},\ }\bibfield  {title} {\bibinfo {title} {Chlorine vacancy in 4h-sic: An nv-like defect with telecom-wavelength emission},\ }\href {https://doi.org/10.1103/PhysRevB.108.224106} {\bibfield  {journal} {\bibinfo  {journal} {Phys. Rev. B}\ }\textbf {\bibinfo {volume} {108}},\ \bibinfo {pages} {224106} (\bibinfo {year} {2023})}\BibitemShut {NoStop}%
\bibitem [{\citenamefont {Jain}\ \emph {et~al.}(2013)\citenamefont {Jain}, \citenamefont {Ong}, \citenamefont {Hautier}, \citenamefont {Chen}, \citenamefont {Richards}, \citenamefont {Dacek}, \citenamefont {Cholia}, \citenamefont {Gunter}, \citenamefont {Skinner}, \citenamefont {Ceder},\ and\ \citenamefont {Persson}}]{10.1063/1.4812323}%
  \BibitemOpen
  \bibfield  {author} {\bibinfo {author} {\bibfnamefont {A.}~\bibnamefont {Jain}}, \bibinfo {author} {\bibfnamefont {S.~P.}\ \bibnamefont {Ong}}, \bibinfo {author} {\bibfnamefont {G.}~\bibnamefont {Hautier}}, \bibinfo {author} {\bibfnamefont {W.}~\bibnamefont {Chen}}, \bibinfo {author} {\bibfnamefont {W.~D.}\ \bibnamefont {Richards}}, \bibinfo {author} {\bibfnamefont {S.}~\bibnamefont {Dacek}}, \bibinfo {author} {\bibfnamefont {S.}~\bibnamefont {Cholia}}, \bibinfo {author} {\bibfnamefont {D.}~\bibnamefont {Gunter}}, \bibinfo {author} {\bibfnamefont {D.}~\bibnamefont {Skinner}}, \bibinfo {author} {\bibfnamefont {G.}~\bibnamefont {Ceder}},\ and\ \bibinfo {author} {\bibfnamefont {K.~A.}\ \bibnamefont {Persson}},\ }\bibfield  {title} {\bibinfo {title} {Commentary: The materials project: A materials genome approach to accelerating materials innovation},\ }\href {https://doi.org/10.1063/1.4812323} {\bibfield  {journal} {\bibinfo  {journal} {APL Materials}\ }\textbf {\bibinfo {volume} {1}},\ \bibinfo {pages}
  {011002} (\bibinfo {year} {2013})},\ \Eprint {https://arxiv.org/abs/https://pubs.aip.org/aip/apm/article-pdf/doi/10.1063/1.4812323/13163869/011002\_1\_online.pdf} {https://pubs.aip.org/aip/apm/article-pdf/doi/10.1063/1.4812323/13163869/011002\_1\_online.pdf} \BibitemShut {NoStop}%
\bibitem [{\citenamefont {Kanai}\ \emph {et~al.}(2022)\citenamefont {Kanai}, \citenamefont {Heremans}, \citenamefont {Seo}, \citenamefont {Wolfowicz}, \citenamefont {Anderson}, \citenamefont {Sullivan}, \citenamefont {Onizhuk}, \citenamefont {Galli}, \citenamefont {Awschalom},\ and\ \citenamefont {Ohno}}]{kanai}%
  \BibitemOpen
  \bibfield  {author} {\bibinfo {author} {\bibfnamefont {S.}~\bibnamefont {Kanai}}, \bibinfo {author} {\bibfnamefont {F.~J.}\ \bibnamefont {Heremans}}, \bibinfo {author} {\bibfnamefont {H.}~\bibnamefont {Seo}}, \bibinfo {author} {\bibfnamefont {G.}~\bibnamefont {Wolfowicz}}, \bibinfo {author} {\bibfnamefont {C.~P.}\ \bibnamefont {Anderson}}, \bibinfo {author} {\bibfnamefont {S.~E.}\ \bibnamefont {Sullivan}}, \bibinfo {author} {\bibfnamefont {M.}~\bibnamefont {Onizhuk}}, \bibinfo {author} {\bibfnamefont {G.}~\bibnamefont {Galli}}, \bibinfo {author} {\bibfnamefont {D.~D.}\ \bibnamefont {Awschalom}},\ and\ \bibinfo {author} {\bibfnamefont {H.}~\bibnamefont {Ohno}},\ }\bibfield  {title} {\bibinfo {title} {Generalized scaling of spin qubit coherence in over 12,000 host materials},\ }\href {https://doi.org/10.1073/pnas.2121808119} {\bibfield  {journal} {\bibinfo  {journal} {Proceedings of the National Academy of Sciences}\ }\textbf {\bibinfo {volume} {119}},\ \bibinfo {pages} {e2121808119} (\bibinfo {year}
  {2022})},\ \Eprint {https://arxiv.org/abs/https://www.pnas.org/doi/pdf/10.1073/pnas.2121808119} {https://www.pnas.org/doi/pdf/10.1073/pnas.2121808119} \BibitemShut {NoStop}%
\bibitem [{\citenamefont {Davidsson}\ \emph {et~al.}(2024{\natexlab{b}})\citenamefont {Davidsson}, \citenamefont {Onizhuk}, \citenamefont {Vorwerk},\ and\ \citenamefont {Galli}}]{davidsson2024discovery}%
  \BibitemOpen
  \bibfield  {author} {\bibinfo {author} {\bibfnamefont {J.}~\bibnamefont {Davidsson}}, \bibinfo {author} {\bibfnamefont {M.}~\bibnamefont {Onizhuk}}, \bibinfo {author} {\bibfnamefont {C.}~\bibnamefont {Vorwerk}},\ and\ \bibinfo {author} {\bibfnamefont {G.}~\bibnamefont {Galli}},\ }\bibfield  {title} {\bibinfo {title} {Discovery of atomic clock-like spin defects in simple oxides from first principles},\ }\href {https://doi.org/10.1038/s41467-024-49057-8} {\bibfield  {journal} {\bibinfo  {journal} {Nature Communications}\ }\textbf {\bibinfo {volume} {15}},\ \bibinfo {pages} {4812} (\bibinfo {year} {2024}{\natexlab{b}})}\BibitemShut {NoStop}%
\bibitem [{\citenamefont {Somjit}\ \emph {et~al.}(2025)\citenamefont {Somjit}, \citenamefont {Davidsson}, \citenamefont {Jin},\ and\ \citenamefont {Galli}}]{somjit2024nvcentermagnesiumoxide}%
  \BibitemOpen
  \bibfield  {author} {\bibinfo {author} {\bibfnamefont {V.}~\bibnamefont {Somjit}}, \bibinfo {author} {\bibfnamefont {J.}~\bibnamefont {Davidsson}}, \bibinfo {author} {\bibfnamefont {Y.}~\bibnamefont {Jin}},\ and\ \bibinfo {author} {\bibfnamefont {G.}~\bibnamefont {Galli}},\ }\bibfield  {title} {\bibinfo {title} {An nv-center in magnesium oxide as a spin qubit for hybrid quantum technologies},\ }\href {https://doi.org/10.1038/s41524-025-01558-w} {\bibfield  {journal} {\bibinfo  {journal} {npj Computational Materials}\ }\textbf {\bibinfo {volume} {11}},\ \bibinfo {pages} {74} (\bibinfo {year} {2025})}\BibitemShut {NoStop}%
\bibitem [{Note2()}]{Note2}%
  \BibitemOpen
  \bibinfo {note} {More defects in CaO due to inclusion of complexes up to fourth nearest neighbors.}\BibitemShut {Stop}%
\bibitem [{\citenamefont {Ferrenti}\ \emph {et~al.}(2020)\citenamefont {Ferrenti}, \citenamefont {de~Leon}, \citenamefont {Thompson},\ and\ \citenamefont {Cava}}]{Ferrenti2020}%
  \BibitemOpen
  \bibfield  {author} {\bibinfo {author} {\bibfnamefont {A.~M.}\ \bibnamefont {Ferrenti}}, \bibinfo {author} {\bibfnamefont {N.~P.}\ \bibnamefont {de~Leon}}, \bibinfo {author} {\bibfnamefont {J.~D.}\ \bibnamefont {Thompson}},\ and\ \bibinfo {author} {\bibfnamefont {R.~J.}\ \bibnamefont {Cava}},\ }\bibfield  {title} {\bibinfo {title} {Identifying candidate hosts for quantum defects via data mining},\ }\href {https://doi.org/10.1038/s41524-020-00391-7} {\bibfield  {journal} {\bibinfo  {journal} {npj Computational Materials}\ }\textbf {\bibinfo {volume} {6}},\ \bibinfo {pages} {126} (\bibinfo {year} {2020})}\BibitemShut {NoStop}%
\bibitem [{\citenamefont {Groppfeldt}\ \emph {et~al.}(2025)\citenamefont {Groppfeldt}, \citenamefont {Davidsson},\ and\ \citenamefont {Armiento}}]{oscarmaster}%
  \BibitemOpen
  \bibfield  {author} {\bibinfo {author} {\bibfnamefont {O.}~\bibnamefont {Groppfeldt}}, \bibinfo {author} {\bibfnamefont {J.}~\bibnamefont {Davidsson}},\ and\ \bibinfo {author} {\bibfnamefont {R.}~\bibnamefont {Armiento}},\ }\bibfield  {title} {\bibinfo {title} {High-throughput exploration of nv-like color centers across host materials},\ }\href {https://arxiv.org/abs/2503.23828} {\bibfield  {journal} {\bibinfo  {journal} {arXiv:2503.23828}\ } (\bibinfo {year} {2025})},\ \Eprint {https://arxiv.org/abs/2503.23828} {arXiv:2503.23828 [cond-mat.mtrl-sci]} \BibitemShut {NoStop}%
\bibitem [{Note3()}]{Note3}%
  \BibitemOpen
  \bibinfo {note} {Ratio of NV-like defects lies somewhere from 1 in 600 (180 NV-like defects to 100,000 processed defects) to 1 in 60 (compared to ca 10,000 stable defects on the defect hull).}\BibitemShut {Stop}%
\bibitem [{\citenamefont {Abbas}\ \emph {et~al.}(2025)\citenamefont {Abbas}, \citenamefont {Bulancea-Lindvall}, \citenamefont {Davidsson}, \citenamefont {Armiento},\ and\ \citenamefont {Abrikosov}}]{abbas2025theoreticalcharacterizationnvlikedefects}%
  \BibitemOpen
  \bibfield  {author} {\bibinfo {author} {\bibfnamefont {G.}~\bibnamefont {Abbas}}, \bibinfo {author} {\bibfnamefont {O.}~\bibnamefont {Bulancea-Lindvall}}, \bibinfo {author} {\bibfnamefont {J.}~\bibnamefont {Davidsson}}, \bibinfo {author} {\bibfnamefont {R.}~\bibnamefont {Armiento}},\ and\ \bibinfo {author} {\bibfnamefont {I.~A.}\ \bibnamefont {Abrikosov}},\ }\bibfield  {title} {\bibinfo {title} {Theoretical characterization of nv-like defects in 4h-sic using adaq with scan and r2scan meta-gga functionals},\ }\href {https://doi.org/10.1063/5.0252129} {\bibfield  {journal} {\bibinfo  {journal} {Applied Physics Letters}\ }\textbf {\bibinfo {volume} {126}},\ \bibinfo {pages} {154001} (\bibinfo {year} {2025})},\ \Eprint {https://arxiv.org/abs/https://pubs.aip.org/aip/apl/article-pdf/doi/10.1063/5.0252129/20489086/154001\_1\_5.0252129.pdf} {https://pubs.aip.org/aip/apl/article-pdf/doi/10.1063/5.0252129/20489086/154001\_1\_5.0252129.pdf} \BibitemShut {NoStop}%
\bibitem [{\citenamefont {Harris}\ \emph {et~al.}(2020)\citenamefont {Harris}, \citenamefont {Ciccarino}, \citenamefont {Flick}, \citenamefont {Englund},\ and\ \citenamefont {Narang}}]{PhysRevB.102.195206}%
  \BibitemOpen
  \bibfield  {author} {\bibinfo {author} {\bibfnamefont {I.}~\bibnamefont {Harris}}, \bibinfo {author} {\bibfnamefont {C.~J.}\ \bibnamefont {Ciccarino}}, \bibinfo {author} {\bibfnamefont {J.}~\bibnamefont {Flick}}, \bibinfo {author} {\bibfnamefont {D.~R.}\ \bibnamefont {Englund}},\ and\ \bibinfo {author} {\bibfnamefont {P.}~\bibnamefont {Narang}},\ }\bibfield  {title} {\bibinfo {title} {Group-iii quantum defects in diamond are stable spin-1 color centers},\ }\href {https://doi.org/10.1103/PhysRevB.102.195206} {\bibfield  {journal} {\bibinfo  {journal} {Phys. Rev. B}\ }\textbf {\bibinfo {volume} {102}},\ \bibinfo {pages} {195206} (\bibinfo {year} {2020})}\BibitemShut {NoStop}%
\bibitem [{\citenamefont {von Bardeleben}\ and\ \citenamefont {Cantin}(2017)}]{Bardeleben_Cantin_2017}%
  \BibitemOpen
  \bibfield  {author} {\bibinfo {author} {\bibfnamefont {H.}~\bibnamefont {von Bardeleben}}\ and\ \bibinfo {author} {\bibfnamefont {J.}~\bibnamefont {Cantin}},\ }\bibfield  {title} {\bibinfo {title} {Nv centers in silicon carbide: from theoretical predictions to experimental observation},\ }\href {https://doi.org/10.1557/mrc.2017.56} {\bibfield  {journal} {\bibinfo  {journal} {MRS Communications}\ }\textbf {\bibinfo {volume} {7}},\ \bibinfo {pages} {591–594} (\bibinfo {year} {2017})}\BibitemShut {NoStop}%
\bibitem [{\citenamefont {Kobayashi}\ \emph {et~al.}(2023)\citenamefont {Kobayashi}, \citenamefont {Shimura},\ and\ \citenamefont {Watanabe}}]{10.1063/5.0169147}%
  \BibitemOpen
  \bibfield  {author} {\bibinfo {author} {\bibfnamefont {T.}~\bibnamefont {Kobayashi}}, \bibinfo {author} {\bibfnamefont {T.}~\bibnamefont {Shimura}},\ and\ \bibinfo {author} {\bibfnamefont {H.}~\bibnamefont {Watanabe}},\ }\bibfield  {title} {\bibinfo {title} {Oxygen-vacancy defect in 4h-sic as a near-infrared emitter: An ab initio study},\ }\href {https://doi.org/10.1063/5.0169147} {\bibfield  {journal} {\bibinfo  {journal} {Journal of Applied Physics}\ }\textbf {\bibinfo {volume} {134}},\ \bibinfo {pages} {145701} (\bibinfo {year} {2023})},\ \Eprint {https://arxiv.org/abs/https://pubs.aip.org/aip/jap/article-pdf/doi/10.1063/5.0169147/18165046/145701\_1\_5.0169147.pdf} {https://pubs.aip.org/aip/jap/article-pdf/doi/10.1063/5.0169147/18165046/145701\_1\_5.0169147.pdf} \BibitemShut {NoStop}%
\bibitem [{\citenamefont {Goss}\ \emph {et~al.}(2005)\citenamefont {Goss}, \citenamefont {Briddon}, \citenamefont {Rayson}, \citenamefont {Sque},\ and\ \citenamefont {Jones}}]{PhysRevB.72.035214}%
  \BibitemOpen
  \bibfield  {author} {\bibinfo {author} {\bibfnamefont {J.~P.}\ \bibnamefont {Goss}}, \bibinfo {author} {\bibfnamefont {P.~R.}\ \bibnamefont {Briddon}}, \bibinfo {author} {\bibfnamefont {M.~J.}\ \bibnamefont {Rayson}}, \bibinfo {author} {\bibfnamefont {S.~J.}\ \bibnamefont {Sque}},\ and\ \bibinfo {author} {\bibfnamefont {R.}~\bibnamefont {Jones}},\ }\bibfield  {title} {\bibinfo {title} {Vacancy-impurity complexes and limitations for implantation doping of diamond},\ }\href {https://doi.org/10.1103/PhysRevB.72.035214} {\bibfield  {journal} {\bibinfo  {journal} {Phys. Rev. B}\ }\textbf {\bibinfo {volume} {72}},\ \bibinfo {pages} {035214} (\bibinfo {year} {2005})}\BibitemShut {NoStop}%
\bibitem [{\citenamefont {Zhang}\ \emph {et~al.}(2014)\citenamefont {Zhang}, \citenamefont {Tang}, \citenamefont {Zhao}, \citenamefont {Cheng}, \citenamefont {Tu}, \citenamefont {Cong}, \citenamefont {Peng}, \citenamefont {Zhu},\ and\ \citenamefont {Chu}}]{10.1063/1.4892654}%
  \BibitemOpen
  \bibfield  {author} {\bibinfo {author} {\bibfnamefont {Y.~G.}\ \bibnamefont {Zhang}}, \bibinfo {author} {\bibfnamefont {Z.}~\bibnamefont {Tang}}, \bibinfo {author} {\bibfnamefont {X.~G.}\ \bibnamefont {Zhao}}, \bibinfo {author} {\bibfnamefont {G.~D.}\ \bibnamefont {Cheng}}, \bibinfo {author} {\bibfnamefont {Y.}~\bibnamefont {Tu}}, \bibinfo {author} {\bibfnamefont {W.~T.}\ \bibnamefont {Cong}}, \bibinfo {author} {\bibfnamefont {W.}~\bibnamefont {Peng}}, \bibinfo {author} {\bibfnamefont {Z.~Q.}\ \bibnamefont {Zhu}},\ and\ \bibinfo {author} {\bibfnamefont {J.~H.}\ \bibnamefont {Chu}},\ }\bibfield  {title} {\bibinfo {title} {A neutral oxygen-vacancy center in diamond: A plausible qubit candidate and its spintronic and electronic properties},\ }\href {https://doi.org/10.1063/1.4892654} {\bibfield  {journal} {\bibinfo  {journal} {Applied Physics Letters}\ }\textbf {\bibinfo {volume} {105}},\ \bibinfo {pages} {052107} (\bibinfo {year} {2014})},\ \Eprint
  {https://arxiv.org/abs/https://pubs.aip.org/aip/apl/article-pdf/doi/10.1063/1.4892654/13300398/052107\_1\_online.pdf} {https://pubs.aip.org/aip/apl/article-pdf/doi/10.1063/1.4892654/13300398/052107\_1\_online.pdf} \BibitemShut {NoStop}%
\bibitem [{\citenamefont {Goss}\ \emph {et~al.}(2024)\citenamefont {Goss}, \citenamefont {Lowery}, \citenamefont {Briddon},\ and\ \citenamefont {Rayson}}]{GOSS2024110811}%
  \BibitemOpen
  \bibfield  {author} {\bibinfo {author} {\bibfnamefont {J.}~\bibnamefont {Goss}}, \bibinfo {author} {\bibfnamefont {R.}~\bibnamefont {Lowery}}, \bibinfo {author} {\bibfnamefont {P.}~\bibnamefont {Briddon}},\ and\ \bibinfo {author} {\bibfnamefont {M.}~\bibnamefont {Rayson}},\ }\bibfield  {title} {\bibinfo {title} {Density functional theory study of al, ga and in impurities in diamond},\ }\href {https://doi.org/https://doi.org/10.1016/j.diamond.2024.110811} {\bibfield  {journal} {\bibinfo  {journal} {Diamond and Related Materials}\ }\textbf {\bibinfo {volume} {142}},\ \bibinfo {pages} {110811} (\bibinfo {year} {2024})}\BibitemShut {NoStop}%
\bibitem [{\citenamefont {Heremans}\ \emph {et~al.}(2016)\citenamefont {Heremans}, \citenamefont {Yale},\ and\ \citenamefont {Awschalom}}]{7478018}%
  \BibitemOpen
  \bibfield  {author} {\bibinfo {author} {\bibfnamefont {F.~J.}\ \bibnamefont {Heremans}}, \bibinfo {author} {\bibfnamefont {C.~G.}\ \bibnamefont {Yale}},\ and\ \bibinfo {author} {\bibfnamefont {D.~D.}\ \bibnamefont {Awschalom}},\ }\bibfield  {title} {\bibinfo {title} {Control of spin defects in wide-bandgap semiconductors for quantum technologies},\ }\href {https://doi.org/10.1109/JPROC.2016.2561274} {\bibfield  {journal} {\bibinfo  {journal} {Proceedings of the IEEE}\ }\textbf {\bibinfo {volume} {104}},\ \bibinfo {pages} {2009} (\bibinfo {year} {2016})}\BibitemShut {NoStop}%
\bibitem [{\citenamefont {Iv{\'a}dy}\ \emph {et~al.}(2019)\citenamefont {Iv{\'a}dy}, \citenamefont {Davidsson}, \citenamefont {Delegan}, \citenamefont {Falk}, \citenamefont {Klimov}, \citenamefont {Whiteley}, \citenamefont {Hruszkewycz}, \citenamefont {Holt}, \citenamefont {Heremans}, \citenamefont {Son}, \citenamefont {Awschalom}, \citenamefont {Abrikosov},\ and\ \citenamefont {Gali}}]{stackingpaper}%
  \BibitemOpen
  \bibfield  {author} {\bibinfo {author} {\bibfnamefont {V.}~\bibnamefont {Iv{\'a}dy}}, \bibinfo {author} {\bibfnamefont {J.}~\bibnamefont {Davidsson}}, \bibinfo {author} {\bibfnamefont {N.}~\bibnamefont {Delegan}}, \bibinfo {author} {\bibfnamefont {A.~L.}\ \bibnamefont {Falk}}, \bibinfo {author} {\bibfnamefont {P.~V.}\ \bibnamefont {Klimov}}, \bibinfo {author} {\bibfnamefont {S.~J.}\ \bibnamefont {Whiteley}}, \bibinfo {author} {\bibfnamefont {S.~O.}\ \bibnamefont {Hruszkewycz}}, \bibinfo {author} {\bibfnamefont {M.~V.}\ \bibnamefont {Holt}}, \bibinfo {author} {\bibfnamefont {F.~J.}\ \bibnamefont {Heremans}}, \bibinfo {author} {\bibfnamefont {N.~T.}\ \bibnamefont {Son}}, \bibinfo {author} {\bibfnamefont {D.~D.}\ \bibnamefont {Awschalom}}, \bibinfo {author} {\bibfnamefont {I.~A.}\ \bibnamefont {Abrikosov}},\ and\ \bibinfo {author} {\bibfnamefont {A.}~\bibnamefont {Gali}},\ }\bibfield  {title} {\bibinfo {title} {Stabilization of point-defect spin qubits by quantum wells},\ }\href
  {https://doi.org/10.1038/s41467-019-13495-6} {\bibfield  {journal} {\bibinfo  {journal} {Nature Communications}\ }\textbf {\bibinfo {volume} {10}},\ \bibinfo {pages} {5607} (\bibinfo {year} {2019})}\BibitemShut {NoStop}%
\bibitem [{\citenamefont {Stenlund}\ \emph {et~al.}(2025)\citenamefont {Stenlund}, \citenamefont {Davidsson}, \citenamefont {Armiento}, \citenamefont {Ivády},\ and\ \citenamefont {Abrikosov}}]{STENLUND2025109468}%
  \BibitemOpen
  \bibfield  {author} {\bibinfo {author} {\bibfnamefont {W.}~\bibnamefont {Stenlund}}, \bibinfo {author} {\bibfnamefont {J.}~\bibnamefont {Davidsson}}, \bibinfo {author} {\bibfnamefont {R.}~\bibnamefont {Armiento}}, \bibinfo {author} {\bibfnamefont {V.}~\bibnamefont {Ivády}},\ and\ \bibinfo {author} {\bibfnamefont {I.~A.}\ \bibnamefont {Abrikosov}},\ }\bibfield  {title} {\bibinfo {title} {Adaq-sym: Automated symmetry analysis of defect orbitals},\ }\href {https://doi.org/https://doi.org/10.1016/j.cpc.2024.109468} {\bibfield  {journal} {\bibinfo  {journal} {Computer Physics Communications}\ }\textbf {\bibinfo {volume} {308}},\ \bibinfo {pages} {109468} (\bibinfo {year} {2025})}\BibitemShut {NoStop}%
\bibitem [{\citenamefont {Shafizadeh}\ \emph {et~al.}(2024)\citenamefont {Shafizadeh}, \citenamefont {Davidsson}, \citenamefont {Ohshima}, \citenamefont {Abrikosov}, \citenamefont {Son},\ and\ \citenamefont {Ivanov}}]{shafizadeh2023selection}%
  \BibitemOpen
  \bibfield  {author} {\bibinfo {author} {\bibfnamefont {D.}~\bibnamefont {Shafizadeh}}, \bibinfo {author} {\bibfnamefont {J.}~\bibnamefont {Davidsson}}, \bibinfo {author} {\bibfnamefont {T.}~\bibnamefont {Ohshima}}, \bibinfo {author} {\bibfnamefont {I.~A.}\ \bibnamefont {Abrikosov}}, \bibinfo {author} {\bibfnamefont {N.~T.}\ \bibnamefont {Son}},\ and\ \bibinfo {author} {\bibfnamefont {I.~G.}\ \bibnamefont {Ivanov}},\ }\bibfield  {title} {\bibinfo {title} {Selection rules in the excitation of the divacancy and the nitrogen-vacancy pair in 4h- and 6h-sic},\ }\href {https://doi.org/10.1103/PhysRevB.109.235203} {\bibfield  {journal} {\bibinfo  {journal} {Phys. Rev. B}\ }\textbf {\bibinfo {volume} {109}},\ \bibinfo {pages} {235203} (\bibinfo {year} {2024})}\BibitemShut {NoStop}%
\end{thebibliography}%

\end{document}